\documentclass{article} 
\usepackage{arxiv,times}
\iclrfinalcopy

\usepackage{amsmath,amsfonts,bm}

\def\eqref#1{equation~\ref{#1}}

\def\1{\bm{1}}

\DeclareMathAlphabet{\mathsfit}{\encodingdefault}{\sfdefault}{m}{sl}
\SetMathAlphabet{\mathsfit}{bold}{\encodingdefault}{\sfdefault}{bx}{n}

\usepackage[T1]{fontenc}
\usepackage{hyperref}
\usepackage{url}
\usepackage{multirow}
\usepackage{graphicx}
\usepackage[table,xcdraw]{xcolor} 
\usepackage{booktabs}
\usepackage{wrapfig}

\usepackage{amsmath}
\usepackage{amssymb}
\usepackage{tabularx}
\usepackage{mathtools}
\usepackage{amsthm}
\usepackage{graphicx}
\usepackage{svg}
\usepackage{pifont}
\usepackage{enumitem}
\usepackage{graphicx}
\usepackage{capt-of}

\usepackage[most,listings]{tcolorbox}
\usepackage{listings}
\usepackage{xcolor}

\tcbuselibrary{breakable,skins,listings}

\definecolor{promptblue}{HTML}{082F4F}
\definecolor{promptbackground}{HTML}{F7F8FA}
\definecolor{promptarrow}{HTML}{E9A800}

\tcbset{
  promptbox/.style={
    enhanced,
    breakable,
    colback=promptbackground,
    colframe=promptblue,
    colbacktitle=promptblue,
    coltitle=white,
    fonttitle=\rmfamily\bfseries\small,
    boxrule=0.6pt,
    arc=1.8mm,
    outer arc=1.8mm,
    lefttitle=2.5mm,
    righttitle=2.5mm,
    toptitle=1.2mm,
    bottomtitle=1.2mm,
    left=3mm,
    right=3mm,
    top=2.5mm,
    bottom=2.5mm,
    before skip=7pt,
    after skip=7pt,
    listing only,
    listing engine=listings,
    listing options={
      basicstyle=\rmfamily\footnotesize,
      numbers=none,
      breaklines=true,
      breakatwhitespace=false,
      columns=fullflexible,
      keepspaces=true,
      showstringspaces=false,
      upquote=true,
      tabsize=2,
      postbreak=\mbox{\textcolor{promptarrow}{$\hookrightarrow$}\space},
      breakindent=0 em,
    },
  }
}

\newcommand{\promptcaption}[2]{%
  \begingroup
  \setlength{\abovecaptionskip}{0pt}%
  \captionof{figure}{#1}%
  \label{#2}%
  \endgroup
}

\newcommand{\xmark}{\textcolor{red}{\ding{55}}}   

\newcommand{\cmark}{\ding{51}} 

\title{CyberClear: A Benchmark for LLM Agent Systems on APT Attack Chain Provenance}

\author{
    Qi Chen\textsuperscript{\rm 1},
    Fushuo Huo\textsuperscript{\rm 1}\thanks{Corresponding author.},
    Hangli Shen\textsuperscript{\rm 1},
    Jingcai Guo\textsuperscript{\rm 2},
    Shuhao Li\textsuperscript{\rm 3},
    Guang Cheng\textsuperscript{\rm 1}
    \\
    \textsuperscript{\rm 1}School of Cyber Science and Engineering, Southeast University \\
    \textsuperscript{\rm 2}The Hong Kong Polytechnic University \\
    \textsuperscript{\rm 3}Zhongguancun Laboratory\\
    \texttt{fushuohuo@seu.edu.cn}
}
\begin{document}

\maketitle

\begin{abstract}
Large language model (LLM) agents have demonstrated promising capabilities in cybersecurity tasks, yet their ability to reconstruct complete Advanced Persistent Threat (APT) attack campaigns from complex security logs remains largely unexplored. 
Existing cybersecurity benchmarks for agents mainly focus on vulnerability discovery, exploitation, and security analysis tasks, leaving the evaluation of attack chain provenance under realistic security logs insufficiently studied.
To address this gap, we introduce CyberClear, a benchmark for evaluating LLM agents and advanced agent systems on APT attack chain provenance from long-context security logs. 
CyberClear covers both single-step attacks and multi-stage attack chains, requiring agents to identify attack evidence, infer attack progression, and generate provenance graphs containing entities, causal relationships, MITRE ATT\&CK techniques, and forensic evidence.
To enable comprehensive evaluation, we develop an evaluation method tailored to APT attack chain provenance. 
Unlike conventional text similarity metrics that focus on surface-level matching, our evaluation examines whether reconstructed graphs preserve the semantics of attack chains across single-step behavior correctness, multi-step behavior identification, temporal and causal consistency, entity and relationship fidelity, and overall attack narrative consistency.
Advanced multi-agent systems powered by state-of-the-art LLMs still struggle on CyberClear, motivating us to propose CyberProvenance, an agent cyber harness designed for multi-agents that augments LLM agents with evidence accumulation, execution-based validation, and feedback-guided refinement mechanisms for reliable attack-chain provenance.
Extensive evaluations on CyberClear demonstrate the effectiveness of CyberProvenance in improving evidence reasoning, execution-grounded validation, and complete APT attack chain reconstruction.
The full benchmark and source code are available at \url{https://cyberclear-bench.github.io/cyberclear/}.
\end{abstract}

\section{Introduction}

Large language model (LLM) agents are taking on increasingly complex tasks through multi-step reasoning, tool use, and interaction with external environments~\citep{yao2022react,schick2023toolformer,wang2025openhands}.
Their capabilities have extended to cybersecurity, including vulnerability discovery, exploitation, and security analysis~\citep{deng2023pentestgpt,fang2024llm}.
As agents assume a greater role in security work, comprehensive benchmarks are needed to evaluate their capabilities under realistic conditions.

Existing cybersecurity benchmarks evaluate agents through classic capture-the-flag (CTF) challenges~\citep{shao2024nyuctf,zhang2025cybench} and vulnerabilities drawn from real software projects~\citep{carlini2025autoadvexbench,zhu2025cve,zhang2026bountybench,lee2025secbench,wang2026cybergym}.
These benchmarks test whether agents can solve security challenges, discover or reproduce vulnerabilities, and repair vulnerable code.
However, these evaluations do not directly assess agents' ability to reason about attack progression and reconstruct complete attack chains.

Advanced persistent threat (APT) attack chain provenance addresses this task by correlating security evidence to recover how an attack campaign unfolded~\citep{li2021threat,zipperle2022provenance,zhang2025survey}.
This task presents three main challenges: (i) identifying relevant evidence from long and noisy security logs; (ii) establishing relationships among processes, files, users, and network activities; and (iii) inferring dependencies and temporal progression across multiple attack stages~\citep{phan2026learning,ZHANG2025107670}.
However, existing evaluations largely focus on attack detection or investigation guided by prior attack clues, and lack a unified setting for assessing whether agents can independently discover dispersed evidence and reconstruct complete attack chains without prior attack knowledge.

\begin{figure*}[h]
    \centering
    \includegraphics[width=\linewidth]{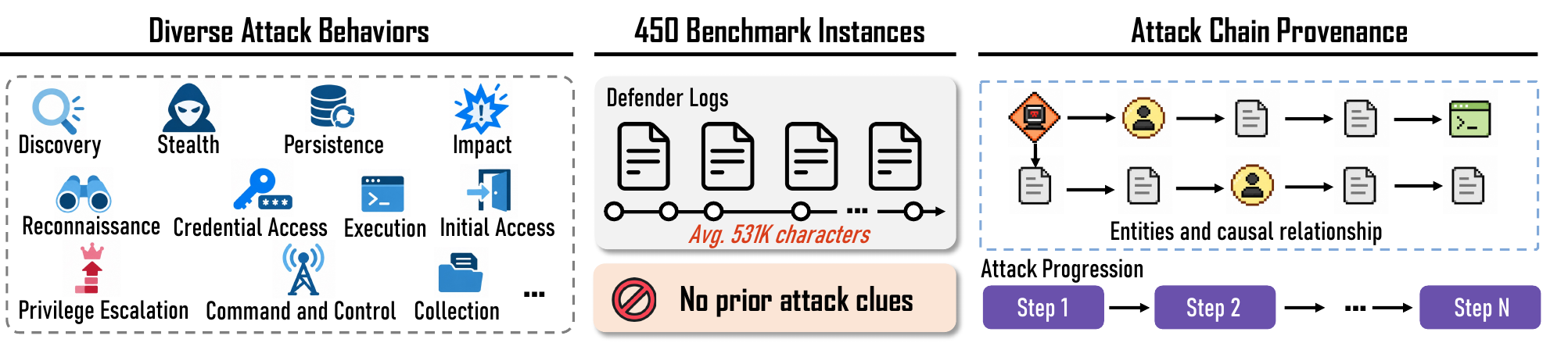}
    \caption{Overview of CyberClear. Given long-context defender logs without prior attack clues, agents are required to identify attack evidence, infer attack progression, and reconstruct complete APT attack chains across multiple attack stages.}
    \label{fig:intro}
\end{figure*}

To address this gap, we introduce CyberClear, a benchmark for evaluating LLM agents and advanced agent systems on APT attack chain provenance. 
As illustrated in Figure~\ref{fig:intro}, CyberClear evaluates whether agents can reconstruct complete APT attack chains from long-context defender observations without prior attack clues.
CyberClear contains 450 instances covering both single-step attacks and multi-stage attack chains, with each instance containing an average of 531K characters of security logs. 
Agents are required to identify attack evidence, infer attack progression, and generate provenance graphs containing attack entities, causal relationships, MITRE ATT\&CK techniques, and forensic evidence.
These elements identify the involved entities, executed actions, and attack strategies, enabling the complete reconstruction of APT attack chains.
Since provenance graphs describing the same attack may differ in representation, we develop a DOT-based LLM judge to evaluate the semantic quality of reconstructed attack chains. 
The proposed evaluation method assesses generated graphs from five perspectives: single-step behavior correctness, multi-step behavior identification, temporal and causal consistency, entity and relationship fidelity, and overall attack narrative consistency, while ignoring irrelevant differences in graph identifiers and visual styles.

Our evaluation reveals that current agents still face significant challenges in long-context evidence reasoning, causal attack progression inference, and complete APT attack chain reconstruction. 
To address these challenges, we propose CyberProvenance, an agent cyber harness for improving the reliability of LLM agents in reconstructing attack chains from long-context security logs in CyberClear.
The harness accumulates evidence across logs, validates predicted attack behaviors in isolated execution environments, and refines inconsistent steps using execution feedback and log evidence.
We evaluate CyberProvenance alongside representative agent frameworks on CyberClear using the proposed metrics. 
Experimental results demonstrate that CyberProvenance effectively performs continuous validation of attack behaviors and reconstructs complete attack chains.

Our main contributions are as follows:
\begin{itemize}
    \item We introduce CyberClear, a comprehensive benchmark for evaluating LLM agents and advanced agent systems on APT attack chain provenance from long-context defender logs without supplied attack clues.
    \item We develop a semantic evaluation method that assesses reconstructed provenance graphs from multiple perspectives, including single-step behavior correctness, multi-step behavior identification, temporal and causal consistency, entity and relationship fidelity, and overall attack consistency.
    \item We propose CyberProvenance, an agent cyber harness designed for multi-agents, which augments LLM agents with evidence accumulation, execution-based validation, and feedback-guided refinement.
    \item We conduct extensive evaluations of representative agent frameworks and LLM backbones on CyberClear, showing that the primary limitations of current agents for APT analysis lie not merely in long context reasoning, but in iterative refinement, evidence consistency, and the effective use of cybersecurity specific verification tools.
\end{itemize}

\section{Related Work}
\paragraph{APT Attack Chain Provenance.}
APT attack chain provenance aims to reconstruct complete attack campaigns by tracing and correlating suspicious activities from large volumes of security logs~\citep{li2021threat,zipperle2022provenance,zhang2025survey}.
Unlike traditional intrusion detection~\citep{du2017deeplog,mirsky2018kitsune,pendlebury2019tesseract} that only identifies isolated malicious events, attack chain provenance focuses on discovering the relationships among entities, behaviors, and attack stages to recover the evolution of multi-step attacks~\citep{phan2026learning,ZHANG2025107670}.
Early methods such as SLEUTH~\citep{hossain2017sleuth}, HOLMES~\citep{milajerdi2019holmes}, and CONAN~\citep{xiong2020conan} rely on predefined expert rules or TTP mappings to identify suspicious activities, but their limited rule coverage often leads to a large number of false positives.
Subsequent graph learning-based approaches~\citep{li2023nodlink,rehman2024flash,cheng2024kairos} leverage provenance graphs to capture complex dependencies among system entities for improved detection, but provide either insufficient contextual information or overly coarse-grained results for attack investigation.
With the rapid development of LLMs, recent studies have increasingly applied LLM-based agents to attack provenance tasks~\citep{mukherjee2025llm,yan2026provagent,zhao2026hunteragent,ock2026distributed}, enabling more flexible reasoning over security logs and attack behaviors.
However, these approaches either identify only entity relationships associated with attack indicators~\citep{mukherjee2025llm,ock2026distributed} or rely on predefined attack clues during evaluation~\citep{yan2026provagent,zhao2026hunteragent}, making them difficult to apply to realistic APT attack chain provenance scenarios where the complete attack process must be reconstructed from security logs.

\begin{table*}[h]
\vspace{-4pt}
    \centering
    \scriptsize
    \setlength{\tabcolsep}{2.5pt}
    \setlength{\belowcaptionskip}{2pt}
    \renewcommand{\arraystretch}{0.9}
    
    \caption{Comparison with existing cybersecurity benchmarks for Agents.}
    \label{tab:benchmark_comparison}

    \begin{tabularx}{\textwidth}{
        @{}
        l
        >{\centering\arraybackslash}X
        >{\centering\arraybackslash}c
        >{\centering\arraybackslash}c
        >{\centering\arraybackslash}c
        >{\centering\arraybackslash}c
        >{\centering\arraybackslash}c
        @{}
    }
    
        \toprule
        \textbf{Benchmark} &
        \textbf{Scope} &
        \textbf{Instance} &
        \textbf{Attack Behaviors} &
        \textbf{ATT\&CK} &
        \textbf{Logs} &
        \textbf{Attack Chain} \\
        \midrule

        NYU CTF Bench~\citep{shao2024nyuctf}
        & CTF & 200 & 6
        & \xmark & \xmark & \xmark \\

        Cybench~\citep{zhang2025cybench}
        & CTF & 40 & 6
        & \xmark & \xmark & \xmark \\

        AutoAdvExBench~\citep{carlini2025autoadvexbench}
        & CTF + Real-world & 75 & 2
        & \xmark & \xmark & \xmark \\

        CVE-Bench~\citep{zhu2025cve}
        & Real-world & 40 & 8
        & \xmark & \xmark & \xmark \\

        BountyBench~\citep{zhang2026bountybench}
        & Real-world & 40 & 36
        & \xmark & \xmark & \xmark \\

        SEC-bench~\citep{lee2025secbench}
        & Real-world & 200 & 16
        & \xmark & \xmark & \xmark \\

        CyberGym~\citep{wang2026cybergym}
        & Real-world & 1,507 & 28
        & \xmark & \xmark & \xmark \\

        DiagChain-Bench~\citep{liu2026}
        & Real-world & 69 & N/R
        & \xmark & \cmark & \cmark \\

        \midrule

        \textbf{CyberClear}
        & \textbf{Real-world}
        & \textbf{450}
        & \textbf{89}
        & \textbf{\cmark}
        & \textbf{\cmark}
        & \textbf{\cmark} \\

        \bottomrule

    \end{tabularx}
    \vspace{-4mm}
\end{table*}

\paragraph{Cybersecurity Benchmarks for Agents.}
The increasing adoption of agents in cybersecurity has motivated the development of benchmarks for evaluating their autonomous security capabilities~\citep{lin2026comparing,sanz2025cybersecurity,carlini2025autoadvexbench}.
As shown in Table~\ref{tab:benchmark_comparison}, existing benchmarks mainly focus on offensive security tasks, including CTF-based problem solving and real-world vulnerability assessment, such as Cybench~\citep{zhang2025cybench}, CVE-Bench~\citep{zhu2025cve}, BountyBench~\citep{zhang2026bountybench}, and CyberGym~\citep{wang2026cybergym}.
While these benchmarks provide realistic environments for evaluating agent capabilities in vulnerability discovery, exploitation, and remediation, they primarily focus on software vulnerability-related tasks, overlooking whether agents can analyze attack evidence from security logs and reason over provenance information to reconstruct complete attack chains.
To address this gap, we introduce a dedicated benchmark, CyberClear, that better emulates realistic APT provenance scenarios featuring long-context security logs and complex multi-stage attacks. 
These characteristics make CyberClear substantially more challenging for existing agents,
as evidenced by our evaluation results in Section~\ref{sec:experiment}.

\section{CyberClear Benchmark}

\subsection{Task Formulation}

We formulate CyberClear as an APT attack chain provenance task that requires agents to reconstruct complete attack processes from large-scale security logs.
Given a collection of defender logs $\boldsymbol{L}=\{L_1,L_2,\ldots,L_n\}$ collected during an attack campaign, the agent performs provenance analysis without access to prior attack clues.
The agent is required to identify attack-related entities, infer their relationships, and reconstruct the complete attack process from dispersed log evidence.
The final output is a provenance graph $\boldsymbol{G}=(E,R)$, where $E$ represents attack-related entities such as processes, accounts, files, services, and network resources, while $R$ captures their interactions and dependencies.
Beyond entity relationships, the graph organizes the reconstructed attack process into an attack narrative timeline, where each step contains its attack behavior, corresponding MITRE ATT\&CK techniques~\citep{jiang2025mitre}, and supporting forensic evidence.

\subsection{Benchmark Construction}

\begin{figure*}[h]
    \centering
    \includegraphics[width=\linewidth]{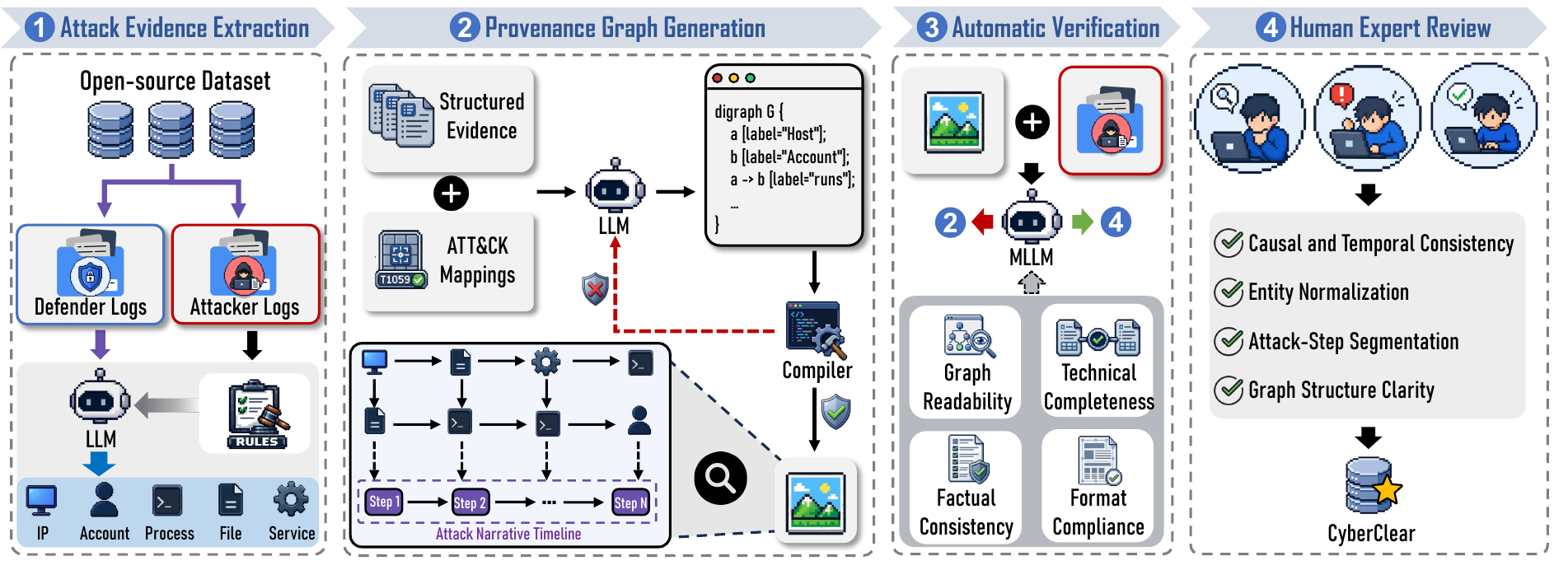}
    \caption{The construction pipeline of CyberClear.}
    \label{fig:bench}
\end{figure*}

Figure~\ref{fig:bench} presents an overview of our four-stage data construction pipeline, including attack evidence extraction, provenance graph generation, automatic verification, and human expert review.
Each sample consists of defender logs collected from a continuous attack campaign and a provenance graph that represents relationships among attack entities and the evolution of multi-stage attacks.

\textbf{Stage I: Attack Evidence Extraction.} 
We collect raw attack logs from publicly available APT datasets, including PROVCON~\citep{yusof2025observations} and CAM-LDS~\citep{landauer2026cam}, which are constructed from real-world attack scenarios.
While designed to support APT behavior analysis and provenance research, these datasets primarily focus on attack detection and security log analysis.
We design attack evidence extraction rules based on attacker logs.
For each sample, we provide Claude Opus 4.8~\citep{anthropic2026claude} with extraction rules to extract attack evidence, including key attack entities, temporal events, entity interactions, step-level attack descriptions, and corresponding MITRE ATT\&CK techniques.
The extracted attack evidence is further organized into structured representations for subsequent processing.

\textbf{Stage II: Provenance Graph Generation.}
We employ Claude Opus 4.8 to generate Graphviz DOT code for each sample based on the extracted attack evidence and MITRE ATT\&CK annotations.
To ensure consistent entity representation in generated DOT graphs, we define unified graph representations that assign predefined node formats to different entity types, including network entities, accounts, processes, files, and services.
The generated DOT code is compiled into visual provenance graphs, and failed compilations are returned to the model for iterative correction.
As shown in Figure~\ref{fig:bench} (2), the resulting provenance graphs preserve entity relationships, and summarize the evolution of multi-stage attacks through attack timelines.

\textbf{Stage III: Automatic Verification.}
The generated graphs only require evidence consistency verification, so we employ GPT-4o~\citep{hurst2024gpt} to automatically verify them using attacker logs as the ground truth.
The verification process consists of four sequential checks: (1) \textbf{Graph readability}, ensuring that the generated graphs are clear, complete, and visually understandable; (2) \textbf{Technical completeness}, verifying that all required ATT\&CK techniques are correctly included; (3) \textbf{Factual consistency}, checking whether the reconstructed entities, relationships, and attack timeline are consistent with the actual attack behaviors; and (4) \textbf{Format compliance}, ensuring that the generated graphs follow predefined visualization and representation requirements.
If any verification criterion fails, the feedback generated by GPT-4o is sent back to step 2 for graph regeneration.

\textbf{Stage IV: Human Expert Review.}
To further ensure dataset reliability, security experts manually review the provenance graphs that pass automatic verification.
The experts evaluate the generated graphs based on predefined quality criteria and filter out low quality samples from open-source datasets, including incomplete attack traces, insufficient evidence coverage, and inconsistent provenance information.
Only samples that satisfy both automatic verification and expert review are included in CyberClear.

\subsection{Data Analysis}

\setlength{\intextsep}{2pt}

\begin{wraptable}{r}{0.45\textwidth}
    \centering
    \scriptsize
    \setlength{\tabcolsep}{3pt}
    \renewcommand{\arraystretch}{0.85}
    \setlength{\belowcaptionskip}{2pt}
    \caption{Statistics of the CyberClear.}
    \label{tab:data_statistics}
    \begin{tabular}{@{}lrrrr@{}}
        \toprule
        \textbf{Statistic} & \textbf{Easy} & \textbf{Medium} & \textbf{Hard} & \textbf{Overall} \\
        \midrule
        \multicolumn{5}{l}{\textbf{Total Samples}} \\
        \quad Sample Count
            & 318 & 122 & 10 & 450 \\
        \midrule
        \multicolumn{5}{l}{\textbf{Attack Complexity}} \\
        \quad Single-step Attack
            & 318 & 0 & 0 & 318 \\
        \quad Multi-step Attack
            & 0 & 122 & 10 & 132 \\
        \midrule
        \multicolumn{5}{l}{\textbf{Input Log Volume (count)}} \\
        \quad Minimum
            & 1 & 1 & 79 & -- \\
        \quad Maximum
            & 12 & 24 & 92 & -- \\
        \quad Average
            & 3.26 & 5.99 & 83.2 & 5.78 \\
        \midrule
        \multicolumn{5}{l}{\textbf{Input Log Length (characters)}} \\
        \quad Minimum
            & 336 & 1167 & 1,222 & -- \\
        \quad Maximum
            & $5.18\mathrm{e}{6}$ & $5.44\mathrm{e}{6}$ & $5.76\mathrm{e}{7}$ & -- \\
        \quad Average
            & $7.07\mathrm{e}{5}$ & $6.69\mathrm{e}{5}$ & $7.80\mathrm{e}{5}$ & $5.31\mathrm{e}{5}$ \\
        \bottomrule
    \end{tabular}
\end{wraptable}

\textbf{Dataset Scale.}
CyberClear contains 450 APT attack chain provenance samples with approximately 2.1GB of data, including 318 easy instances, 122 medium instances, and 10 hard instances categorized by attack complexity.
Table~\ref{tab:data_statistics} summarizes the dataset statistics, including sample distribution, attack complexity, input log volume, and log length.
The easy subset consists entirely of single-step attacks, while medium and hard subsets contain multi-step attack chains requiring models to reason over dependencies among multiple attack stages.
Compared with easy and medium instances, hard samples involve substantially larger input log volumes, with an average of 83.2 log files per sample and longer contexts containing approximately $7.80\times10^{5}$ characters, providing a more challenging setting for evaluating APT attack chain provenance capabilities.

\begin{wrapfigure}{r}{0.45\textwidth}
    \centering
    \includegraphics[width=\linewidth]{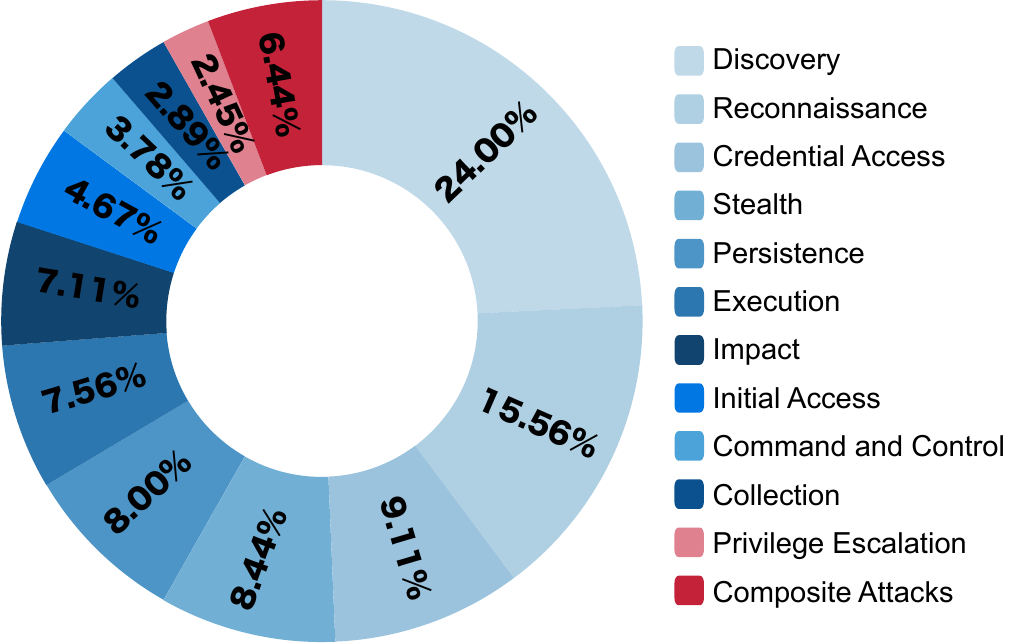}
    \vspace{-14pt}
    \caption{Distribution of attack behaviors.}
    \label{fig:attack_categories}
\end{wrapfigure}

\textbf{Attack Behaviors.}
Figure~\ref{fig:attack_categories} shows the distribution of attack behaviors in CyberClear. 
The benchmark covers 89 distinct attack behaviors spanning Credential Access, Defense Evasion, Persistence, Execution, Impact, and other attack categories. 
Discovery and Reconnaissance are the two most frequent categories, accounting for 24.00\% and 15.56\% of all instances, respectively.
Among them, Composite Attacks are constructed by combining multiple attack behaviors, accounting for 6.44\% of all instances.
This diverse distribution provides broad coverage of APT behaviors across different attack stages and enables comprehensive evaluation of attack chain provenance capabilities.

\subsection{Evaluation Metrics}
\setlength{\intextsep}{-2pt}

\textbf{Code-Level Evaluation.}
The generated DOT code is evaluated from both textual similarity and functional correctness perspectives.
BLEU and ROUGE-L are used to measure the similarity between generated and reference DOT code, capturing token-level overlap and sequence-level consistency, respectively.
CodeBLEU is further adopted to evaluate structural consistency by considering code syntax and semantic information beyond surface-level matching.
Pass@1 is used to assess whether the generated DOT code can be successfully compiled, reflecting practical generation reliability.

\textbf{Semantic Evaluation via LLM-as-a-Judge.}
To complement lexical matching and code structure metrics, we develop a DOT-based LLM judge to evaluate the semantic quality of reconstructed attack chains.
The judge directly compares reference and predicted DOT code while ignoring irrelevant differences in node identifiers, statement order, graph layouts, and visual styles.
We use GLM-5.2~\citep{zeng2026glm} as the judge model to perform evidence-based semantic comparison between reference and predicted provenance graphs.
The semantic evaluation framework assesses reconstructed provenance graphs from five perspectives.
Each dimension is scored from 0 to 1, and the reported scores are averaged over all samples.
Detailed illustrations are in Appendix~\ref{app:Subjective Metrics}.
\begin{itemize}[leftmargin=*, itemsep=3pt, topsep=0pt]
    \item \textbf{Single-Step Judge.} 
    Evaluates whether individual attack behaviors are correctly reconstructed by comparing the actor, core action, target, and overall attack meaning.
    
    \item \textbf{Loose Judge.}
    Measures whether multi-step attack behaviors are correctly identified without considering temporal ordering among attack steps.
    
    \item \textbf{Strict Judge.}
    Evaluates whether the reconstructed attack chain preserves the temporal progression and causal dependencies among different attack stages.
    
    \item \textbf{Detail Judge.}
    Measures the consistency of graph-level elements, including entities, directed relationships, and attributes between reference and predicted provenance graphs.
    
    \item \textbf{Scenario Judge.}
    Assesses whether the reconstructed graph conveys the same overall attack process, target, method, progression, and outcome as the reference attack campaign.
\end{itemize}
\vspace{3mm}
\begin{figure*}[h]
    \centering
    \includegraphics[width=\linewidth]{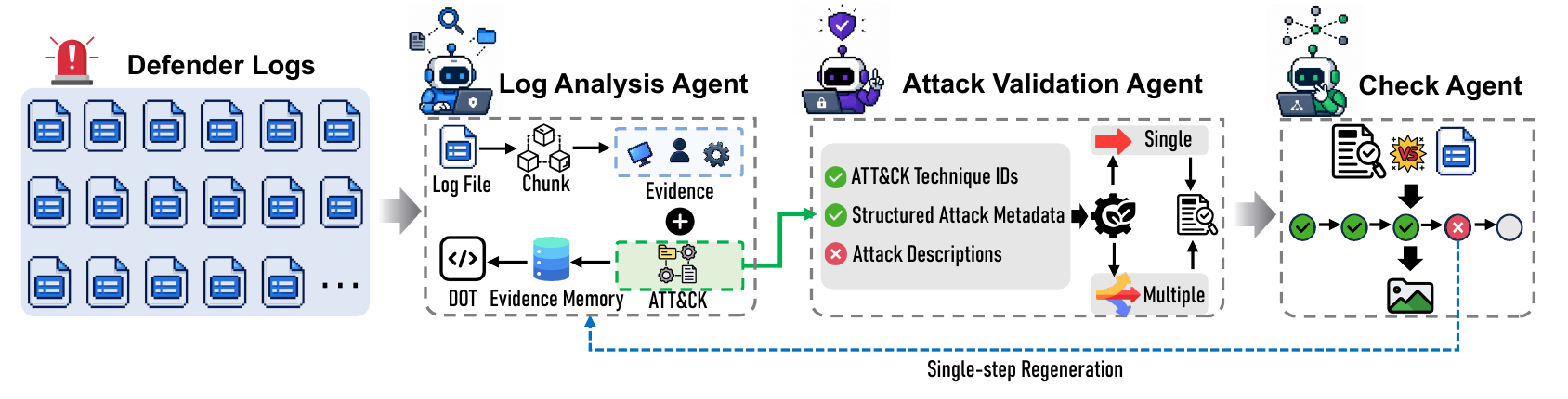}
    \caption{
    Overview of CyberProvenance, an agent cyber harness for reliable attack-chain provenance.
    Designed for multi-agent collaboration, the harness assigns specialized capabilities to different agents, enabling evidence accumulation, execution-based validation, and feedback-guided refinement.
    }
    \label{fig:method}
\end{figure*}
\vspace{3mm}
\section{Methodology}
We present CyberProvenance, an agent cyber harness designed for multi-agent collaboration to improve the reliability of LLM-based agents.
The harness assigns specialized capabilities to different agent roles, including Log Analysis Agent, Attack Validation Agent, and Check Agent, which collaboratively perform evidence accumulation, execution-based validation, and iterative provenance graph refinement, as illustrated in Figure~\ref{fig:method}.



\textbf{Log Analysis Agent.}
To identify and accumulate attack evidence from large volumes of logs and establish an initial provenance structure, the Log Analysis Agent is responsible for evidence accumulation in the CyberProvenance harness.
It receives defender security logs $\boldsymbol{L}=\{L_1,\ldots,L_n\}$ and progressively analyzes logs to identify attack related evidence, including temporal events, processes, commands, network activities, and potential MITRE ATT\&CK techniques.
To handle long context logs, each log file $L_i$ is divided into multiple chunks $\{C_i^1,\ldots,C_i^{K_i}\}$, and the extracted evidence from each chunk is continuously accumulated into an Evidence Memory $\boldsymbol{M}$.
Based on the accumulated evidence, the agent performs global reasoning to generate an initial provenance hypothesis:
\[
\boldsymbol{H}_{0}=\mathcal{F}(\boldsymbol{M})=(\boldsymbol{A},\boldsymbol{S},\boldsymbol{T},\boldsymbol{G}_{0}),
\]
where $\boldsymbol{A}$ denotes the attack scenario, $\boldsymbol{S}$ represents attack stages and step dependencies, $\boldsymbol{T}$ denotes ATT\&CK technique mappings, and $\boldsymbol{G}_{0}$ represents the reconstructed provenance graph.
The reconstructed provenance graph $\boldsymbol{G_0}$ is finally serialized into structured DOT code, providing an explicit graph representation for subsequent attack validation.

\textbf{Attack Validation Agent.}
The Attack Validation Agent provides execution-grounded validation by verifying the executability and reliability of reconstructed attack steps through real environment reproduction.
This reduces the risk of relying on plausible but unverified attack inferences.
Given the predicted ATT\&CK techniques $\boldsymbol{T}=\{\tau_1,\ldots,\tau_m\}$ and structured attack steps generated by the Log Analysis Agent, the agent retrieves the corresponding execution specifications for each technique:
\[
q_i=\mathcal{R}(\tau_i)=(p_i,x_i,r_i,d_i),
\]
where $p_i$, $x_i$, $r_i$, and $d_i$ represent the required platform, executor, privilege, and dependency information of technique $\tau_i$, respectively.
Using these specifications, the agent selects appropriate execution strategies, where independent techniques are reproduced using Atomic Red Team~\citep{atomicredteam} and dependency aware multi step attack chains are orchestrated through CALDERA~\citep{apachecaldera} in isolated environments built with Ludus~\citep{ludus}.
The generated execution traces collected by Sysmon~\citep{sysmon}, Linux Audit~\citep{linuxaudit}, and Zeek~\citep{zeek} are used to produce structured validation reports, indicating whether each reconstructed attack step can be successfully reproduced.

\textbf{Check Agent.}
The Check Agent closes the refinement loop of the harness by integrating the reconstructed provenance graph and validation feedback to assess reliability through graph structures, attack-step dependencies, and ATT\&CK technique mappings.
When all attack steps are successfully reproduced and consistent with the collected evidence, the current provenance graph is preserved as the final result. 
Otherwise, unreliable steps are refined while preserving validated nodes, and the updated hypothesis is revalidated to form an iterative prediction, validation, and refinement loop.

\begin{table*}[h]
\setlength{\belowcaptionskip}{5pt}
\caption{Performance comparison on CyberClear across agent frameworks and LLM backbones using code-level and semantic metrics. Blue cells indicate the best result for each metric.}
\centering
\small
\setlength{\tabcolsep}{1.2pt}
\renewcommand{\arraystretch}{1.2}
\resizebox{\textwidth}{!}{
\begin{tabular}{lc|cccc|ccccc}
\toprule

\multirow{2}{*}{\textbf{Methods}}
& \multirow{2}{*}{}
& \multicolumn{4}{c|}{\textbf{Code-Level Metrics}}
& \multicolumn{5}{c}{\textbf{Semantic Metrics}}
\\

\cmidrule(lr){3-6}
\cmidrule(lr){7-11}

& 
& BLEU 
& ROUGE-L 
& CodeBLEU 
& Pass@1
& Single 
& Loose 
& Strict 
& Detail 
& Scenario
\\

\midrule

\multirow{3}{*}{OpenClaw (Single)}
& \includegraphics[height=0.3cm]{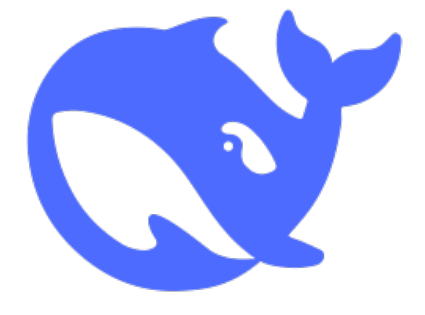}
& 0.3317 & 0.4863 & 0.3795 & 0.8222
& 0.6794 & 0.6006 & 0.4387 & 0.4051 & 0.4805 \\

& \includegraphics[height=0.33cm]{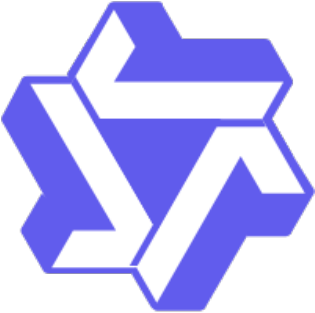}
& 0.3556 & 0.5117 & 0.3868 & 0.7200
& 0.7812 & 0.7125 & 0.5490 & 0.4898 & 0.5823 \\

& \includegraphics[height=0.33cm]{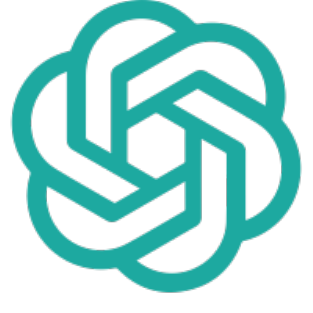}
& 0.3692 & 0.5297 & 0.3887 & 0.8467
& 0.7726 & 0.7214 & 0.5008 & 0.4536 & 0.5862 \\

\midrule

\multirow{3}{*}{OpenClaw (Multiple)}
& \includegraphics[height=0.3cm]{img/deepseek.pdf}
& 0.4183 & 0.5654 & 0.5648 & 0.9200
& 0.5513 & 0.4749 & 0.3580 & 0.3301 & 0.3679 \\

& \includegraphics[height=0.33cm]{img/qwen.pdf}
& 0.4422 & 0.5908 & 0.5721 & 0.8178
& 0.7160 & 0.6629 & 0.5034 & 0.4090 & 0.5847 \\

& \includegraphics[height=0.33cm]{img/gpt.pdf}
& 0.4516 & 0.5848 & 0.5643 & 0.9333
& 0.7136 & 0.6684 & 0.5178 & 0.4053 & 0.5821 \\

\midrule

\multirow{3}{*}{AgentVerse}
& \includegraphics[height=0.3cm]{img/deepseek.pdf}
& 0.3664 & 0.5446 & 0.6074 & 0.9400
& 0.5080 & 0.4337 & 0.3273 & 0.3039 & 0.3269 \\

& \includegraphics[height=0.33cm]{img/qwen.pdf}
& 0.3903 & 0.5700 & 0.6147 & 0.8378
& 0.7304 & 0.6743 & 0.5150 & 0.4248 & 0.5886 \\

& \includegraphics[height=0.33cm]{img/gpt.pdf}
& 0.4015 & 0.5942 & 0.6027 & 0.9422
& 0.7113 & 0.6897 & 0.5011 & 0.4179 & 0.6135 \\

\midrule

\multirow{3}{*}{LLM-Debate}
& \includegraphics[height=0.3cm]{img/deepseek.pdf}
& 0.4089 & 0.5278 & 0.6502 & 0.8644
& 0.6135 & 0.5344 & 0.4089 & 0.3758 & 0.4334 \\

& \includegraphics[height=0.33cm]{img/qwen.pdf}
& 0.4328 & 0.5532 & 0.6575 & 0.7622
& 0.7622 & 0.6853 & 0.5368 & 0.4581 & 0.6061 \\

& \includegraphics[height=0.33cm]{img/gpt.pdf}
& 0.4413 & 0.5789 & 0.6494 & 0.8844
& 0.7436 & 0.7082 & 0.5216 & 0.4704 & 0.6047 \\

\midrule

\multirow{3}{*}{MultiPersona}
& \includegraphics[height=0.3cm]{img/deepseek.pdf}
& 0.3775 & 0.5570 & 0.4217 & 0.9289
& 0.6843 & 0.6033 & 0.4418 & 0.4037 & 0.4842 \\

& \includegraphics[height=0.33cm]{img/qwen.pdf}
& 0.4014 & 0.5824 & 0.4290 & 0.8267
& 0.7800 & 0.7107 & 0.5519 & 0.4825 & 0.5961 \\

& \includegraphics[height=0.33cm]{img/gpt.pdf}
& 0.4136 & 0.5314 & 0.4247 & 0.9422
& 0.7131 & 0.7198 & 0.5289 & 0.5002 & 0.5235 \\

\midrule

\multirow{3}{*}{LLM-Blender}
& \includegraphics[height=0.3cm]{img/deepseek.pdf}
& 0.4196 & 0.5495 & 0.5761 & 0.8733
& 0.5632 & 0.4869 & 0.3660 & 0.3384 & 0.3789 \\

& \includegraphics[height=0.33cm]{img/qwen.pdf}
& 0.4435 & 0.5749 & 0.5834 & 0.7711
& 0.7216 & 0.6668 & 0.5077 & 0.4151 & 0.5864 \\

& \includegraphics[height=0.33cm]{img/gpt.pdf}
& 0.4519 & 0.5976 & 0.6112 & 0.8956
& 0.7317 & 0.6761 & 0.5248 & 0.4465 & 0.5969 \\

\midrule

\multirow{3}{*}{DyLAN}
& \includegraphics[height=0.3cm]{img/deepseek.pdf}
& 0.3497 & 0.5241 & 0.5099 & 0.8844
& 0.8454 & 0.7514 & 0.5421 & 0.4769 & 0.6223 \\

& \includegraphics[height=0.33cm]{img/qwen.pdf}
& 0.3808 & 0.5528 & 0.5491 & 0.8667
& 0.8511 & \cellcolor{blue!10}\textbf{0.7705} & 0.6224 & 0.5433 & 0.6545 \\

& \includegraphics[height=0.33cm]{img/gpt.pdf}
& 0.3884 & 0.5258 & 0.5375 & 0.8711
& 0.8350 & 0.7502 & 0.5592 & 0.5214 & 0.6771 \\

\midrule

\multirow{3}{*}{MacNet}
& \includegraphics[height=0.3cm]{img/deepseek.pdf}
& 0.3633 & 0.4728 & 0.5326 & 0.9267
& 0.6186 & 0.5590 & 0.3928 & 0.3977 & 0.4241 \\

& \includegraphics[height=0.33cm]{img/qwen.pdf}
& 0.3630 & 0.4784 & 0.5422 & 0.7889
& 0.7708 & 0.7129 & 0.5226 & 0.5090 & 0.5061 \\

& \includegraphics[height=0.33cm]{img/gpt.pdf}
& 0.3389 & 0.4515 & 0.5086 & 0.8088
& 0.7375 & 0.6809 & 0.4758 & 0.5070 & 0.5022 \\

\midrule

\multirow{3}{*}{CAMEL}
& \includegraphics[height=0.3cm]{img/deepseek.pdf}
& 0.3577 & 0.5647 & 0.5629 & 0.9222
& 0.3754 & 0.3048 & 0.2371 & 0.2217 & 0.2043 \\

& \includegraphics[height=0.33cm]{img/qwen.pdf}
& 0.4057 & 0.5869 & 0.5790 & 0.7978
& 0.6710 & 0.6322 & 0.4673 & 0.3618 & 0.5656 \\

& \includegraphics[height=0.33cm]{img/gpt.pdf}
& 0.3864 & 0.5781 & 0.4201 & 0.9377
& 0.5371 & 0.4557 & 0.2955 & 0.3025 & 0.4117 \\

\midrule

\multirow{3}{*}{AgentScope-V2}
& \includegraphics[height=0.3cm]{img/deepseek.pdf}
& 0.4921 & 0.6477 & 0.6805 & 0.8622
& 0.8412 & 0.7373 & 0.6291 & 0.5485 & 0.7021 \\

& \includegraphics[height=0.33cm]{img/qwen.pdf}
& 0.4977 & 0.6523 & 0.6854 & 0.7889
& 0.8343 & 0.6786 & 0.5864 & 0.5169 & 0.6936 \\

& \includegraphics[height=0.33cm]{img/gpt.pdf}
& \cellcolor{blue!10}\textbf{0.5035} & 0.6334 & \cellcolor{blue!10}\textbf{0.7240} & 0.9866
& 0.8525 & 0.6597 & 0.6106 & 0.5710 & 0.7115 \\

\midrule

\multirow{3}{*}{CyberProvenance (Ours)}
& \includegraphics[height=0.3cm]{img/deepseek.pdf}
& 0.4971 & 0.6873 & 0.6867 & \cellcolor{blue!10}\textbf{0.9933}
& 0.8627 & 0.7325 & 0.6340 & 0.5981 & 0.6710 \\

& \includegraphics[height=0.33cm]{img/qwen.pdf}
& 0.4810 & 0.6827 & 0.7052 & 0.8911
& 0.8764 & 0.7027 & \cellcolor{blue!10}\textbf{0.6477}
& 0.5776 & 0.7087 \\

& \includegraphics[height=0.33cm]{img/gpt.pdf}
& 0.4952 & \cellcolor{blue!10}\textbf{0.6968} & 0.7016 & 0.9333
& \cellcolor{blue!10}\textbf{0.8897} & 0.7438 & 0.6337 & \cellcolor{blue!10}\textbf{0.6124} & \cellcolor{blue!10}\textbf{0.7263} \\

\bottomrule

\end{tabular}
}
\label{tab:graph_evaluation}
\vspace{-6mm}
\end{table*}

\section{Experiments}
\label{sec:experiment}

\subsection{Experimental Setup}

\textbf{LLMs.}
We evaluate \includegraphics[height=0.32cm]{img/deepseek.pdf}DeepSeek-V4-Flash~\citep{xu2026deepseek}, \includegraphics[height=0.32cm]{img/qwen.pdf}Qwen3.8-Max~\citep{qwen38}, and \includegraphics[height=0.32cm]{img/gpt.pdf}GPT-5.6-Terra~\citep{openai2026gpt56} in the main experiments. 
These models exhibit different reasoning and generation characteristics, allowing us to examine the performance of agent frameworks across different LLM backbones. 
Additional results with \includegraphics[height=0.32cm]{img/gpt.pdf}GPT-4o-mini~\citep{hurst2024gpt} and \includegraphics[height=0.32cm]{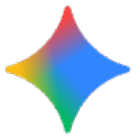}Gemini-3.5-Flash-Lite~\citep{deepmind2026gemini35flash} are reported in Appendix~\ref{app:other_llm}.

\textbf{Methods.}
We compare CyberProvenance against two categories of baselines:
(1) \textbf{Single-Agent Methods} including OpenClaw (Single-Agent)~\citep{openclaw2026};
(2) \textbf{Multi-Agent Systems} including OpenClaw (Multi-Agent), AgentVerse~\citep{chen2024agentverse}, LLM-Debate~\citep{du2023improving}, MultiPersona~\citep{wang2024unleashing}, LLM-Blender~\citep{jiang2023llm}, DyLAN~\citep{liu2024dylan}, MacNet~\citep{qian2025scaling}, CAMEL~\citep{li2023camel}, and AgentScope-V2~\citep{agentscope,agentscope_v1}.
Additional details are provided in Appendix~\ref{app:baselines}.

\vspace{2mm}
\begin{figure*}[h]
    \centering
    \includegraphics[width=\linewidth]{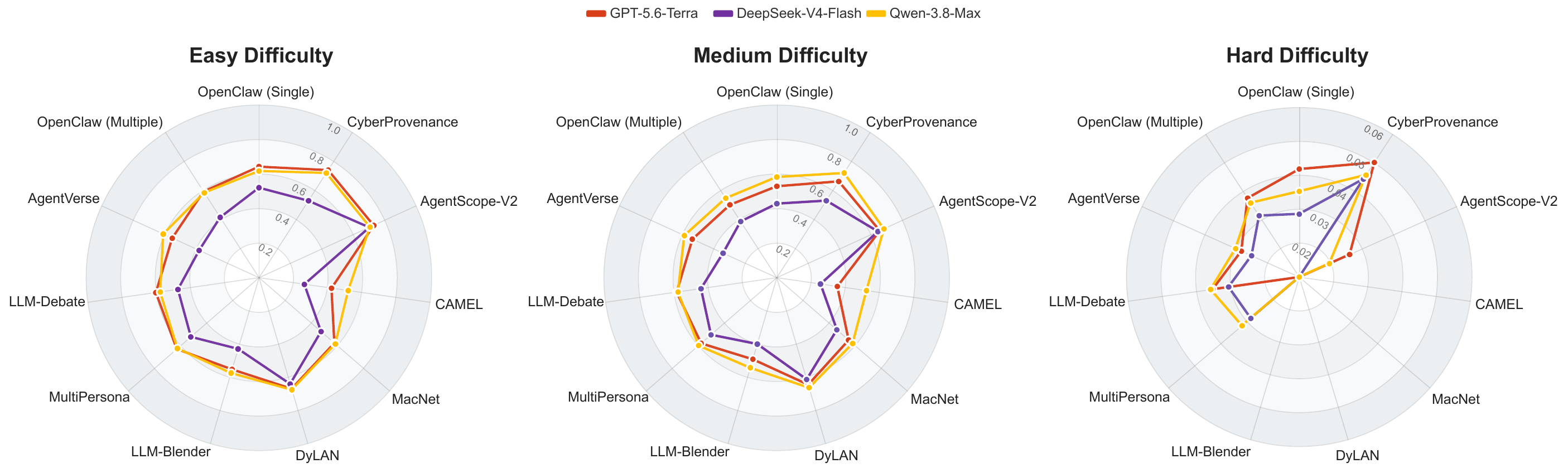}
    \caption{Average Semantic Metrics scores of different agent frameworks across the Easy, Medium, and Hard subsets of CyberClear.}
    \label{fig:radar}
    \vspace{2mm}
\end{figure*}

\subsection{Main Results}

\textbf{Code-Level Metrics.}
Table~\ref{tab:graph_evaluation} shows that existing agent frameworks can generate valid provenance graphs, but their performance varies markedly across methods and LLM backbones, indicating that stable attack-graph reconstruction remains difficult.
AgentScope-V2 achieves the highest BLEU and CodeBLEU scores of 0.5035 and 0.7240, while CyberProvenance obtains the best ROUGE-L score of 0.6968 and Pass@1 score of 0.9933.
Compared with general-purpose agent frameworks, CyberProvenance maintains consistently strong code-level performance across multiple metrics.
These results suggest that iterative refinement and cybersecurity-specific validation tools contribute to more reliable provenance graph reconstruction.

\textbf{Semantic Metrics.}
The challenge is more evident under semantic metrics, where most existing frameworks struggle to preserve complete attack progression, entity dependencies, and fine-grained evidence.
Although DyLAN achieves the best Loose score of 0.7705, its advantage does not extend consistently to stricter semantic criteria.
CyberProvenance achieves the highest Single, Strict, Detail, and Scenario scores of 0.8897, 0.6477, 0.6124, and 0.7263, respectively.
The remaining gap on Strict and Detail further indicates that complete APT attack-chain reconstruction is still challenging for current agent systems.

\vspace{-1mm}
\subsection{Performance Analysis across Difficulty Levels}


We evaluate different agent frameworks across the Easy, Medium, and Hard subsets of CyberClear using the average Semantic Metrics score over Single, Loose, Strict, Detail, and Scenario. 
As shown in Figure~\ref{fig:radar}, all frameworks exhibit consistent performance degradation as the number of input log files increases rapidly. 
This trend indicates that larger log volumes make it increasingly difficult to identify relevant evidence and maintain coherent attack progression during provenance. 
These results demonstrate that CyberClear effectively captures the increasing difficulty of APT attack chain reconstruction across different log scales.

\vspace{3mm}
\begin{figure*}[h]
    \centering
    \includegraphics[width=\linewidth]{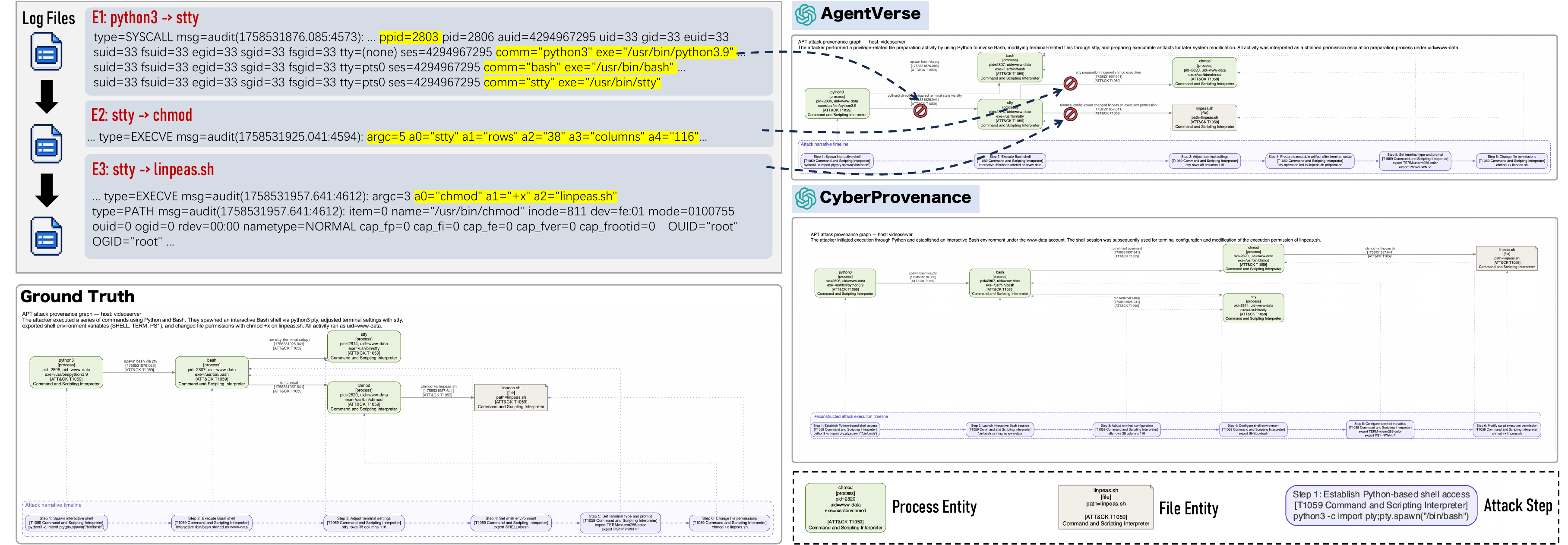}
    \caption{Case study comparing AgentVerse and CyberProvenance on attack provenance using GPT-5.6-Terra. CyberProvenance corrects unsupported causal relations by aligning the reconstructed graph with execution evidence.}
    \label{fig:case}
\end{figure*}

\vspace{3mm}
\subsection{Performance Analysis Across Attack Stage Levels}

\setlength{\intextsep}{-3pt}
\begin{wrapfigure}{r}{0.45\textwidth}
    \centering
    \includegraphics[width=0.88\linewidth]{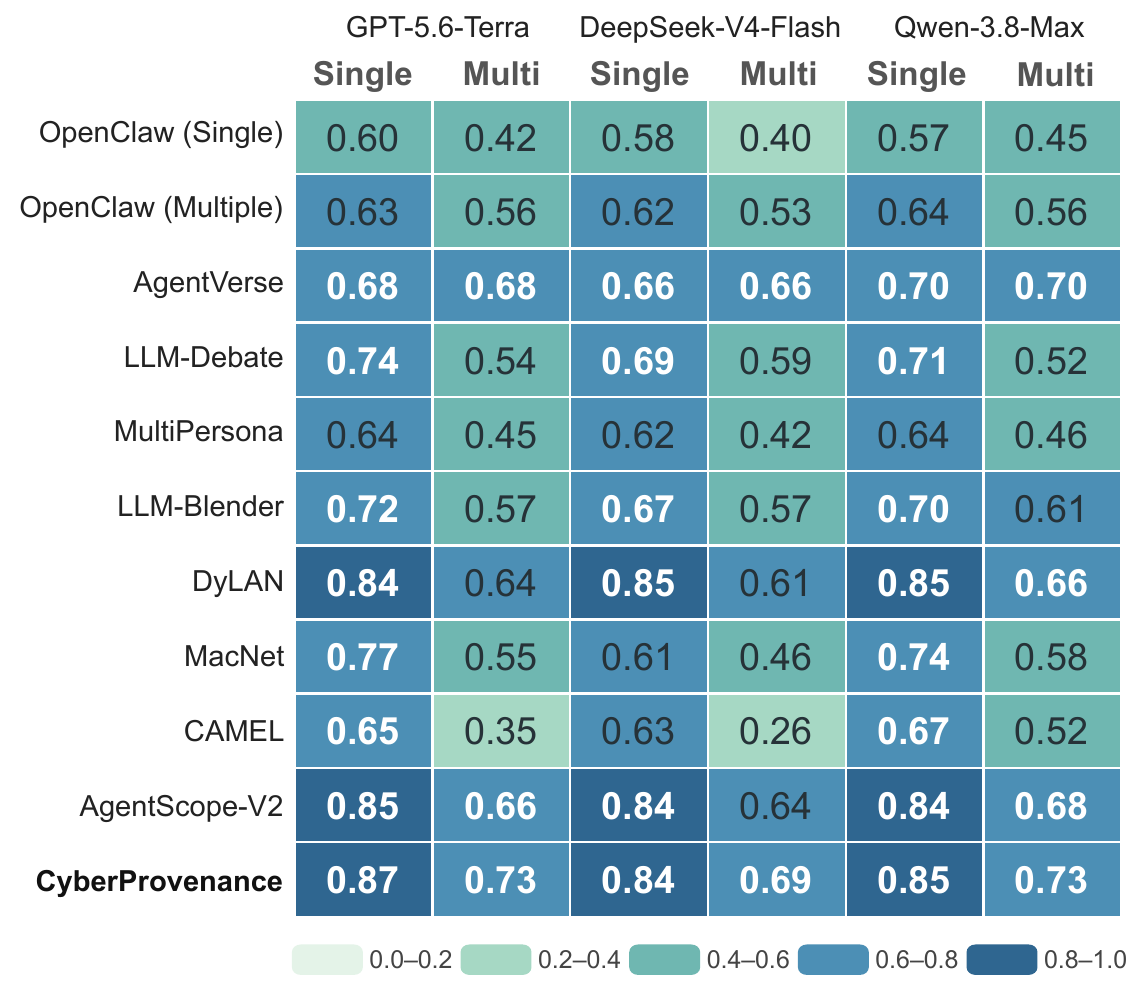}
    \vspace{-16pt}
    \caption{Average Semantic Metrics scores across single-stage and multi-stage attacks.}
    \label{fig:attack_stage}
\end{wrapfigure}

Attack stage complexity exposes a clear performance gap in current agent frameworks. 
As shown in Figure~\ref{fig:attack_stage}, we report the average Semantic Metrics score across Single, Loose, Strict, Detail, and Scenario, where nearly all methods perform substantially worse on multi-stage attacks than on single-stage attacks across the three LLM backbones. 
The decline is especially pronounced for CAMEL, MacNet, and LLM-Debate, whereas CyberProvenance remains more robust and consistently achieves stronger multi-stage performance. 
These results highlight the importance of iterative validation and evidence-guided refinement for maintaining coherent attack chain reconstruction as attack complexity increases.

\subsection{Case Study}


Figure~\ref{fig:case} presents a representative example using GPT-5.6-Terra, where the ground-truth graph shows that \texttt{python3} launches \texttt{bash}, which subsequently invokes both \texttt{stty} and \texttt{chmod}, while \texttt{chmod} modifies \texttt{linpeas.sh}. 
AgentVerse incorrectly associates \texttt{python3} directly with \texttt{stty}, although the audit log shows that \texttt{stty} is spawned by the Bash process rather than Python. 
It further introduces dependencies from \texttt{stty} to \texttt{chmod} and \texttt{linpeas.sh}, even though \texttt{chmod} is independently invoked by Bash and the permission change on \texttt{linpeas.sh} is directly caused by \texttt{chmod +x}. 
In contrast, CyberProvenance removes these unsupported dependencies and reconstructs the causal structure more consistently with the ground truth.
More cases in the Appendix~\ref{app:case}.

\section{Conclusion}
In this work, we introduce CyberClear, a benchmark for evaluating LLM agents on APT attack chain provenance from long-context security logs.
CyberClear covers both single-step and multi-stage attacks and provides a semantic evaluation method that assesses attack behaviors, temporal and causal consistency, entity relationships, and overall attack narrative consistency. 
Extensive evaluations show that even advanced agent systems still face substantial challenges in evidence reasoning and complete attack chain reconstruction. 
To address these limitations, we propose CyberProvenance, demonstrates the effectiveness of an agent cyber harness that combines evidence accumulation, execution-based validation, and feedback-guided refinement for reliable APT attack chain reconstruction.
We hope CyberClear and CyberProvenance will provide a useful foundation for developing more capable and trustworthy agents for complex cybersecurity analysis.


\subsection*{AI use statement}



In this work, we used generative AI tools to assist with refining the research methodology and experimental design, implementing and debugging parts of the experimental code, and interpreting intermediate experimental results. 
We also used generative AI tools for literature search and summarization, language polishing, improving the organization and readability of the manuscript, and assisting with the preparation of figures, tables, and captions. 
All AI-assisted code and experimental procedures were manually reviewed and tested by the authors, and the reported results were checked against the corresponding execution outputs and evaluation scripts. 
Literature-related suggestions were verified against the original papers, and all AI-assisted text, figures, and technical claims were reviewed and revised by the authors before inclusion. 
We take full responsibility for the final content of this work, including all text, claims, code, experiments, and artifacts produced with the aid of generative AI.

\subsection*{Ethics statement}


This work studies defensive cybersecurity analysis and APT attack provenance. 
CyberClear is constructed from publicly available security datasets and does not involve human subjects or private user data collection. 
All execution-based validation is conducted only in authorized and isolated environments, with private network ranges and safeguards that prevent destructive or unauthorized actions against external systems. 
While the studied techniques are inherently dual-use, the benchmark and framework are intended for defensive analysis, provenance, and evaluation of security agents, and we will clearly document the intended use and safety constraints of the released artifacts.

\subsection*{Reproducibility statement}


We have taken several steps to support the reproducibility of this work. 
The main paper describes the benchmark construction, evaluation protocol, and overall experimental setup, while the appendix provides the complete configuration details for all evaluated agent frameworks. 
Additional implementation details, data processing procedures, prompts, and evaluation methods are included in the appendix. 
These materials are intended to provide the information necessary to reproduce the reported experiments and results.


\bibliography{iclr2027_conference}
\bibliographystyle{arxiv}

\newpage

\appendix




\section*{Appendix Overview}

This appendix provides supplementary experiments, analyses, and case studies, organized as follows:

\begin{itemize}[leftmargin=*, itemsep=3pt, topsep=0pt]

    \item \textbf{Appendix~\ref{app:additional_results}: Additional Experiments.}
    We report additional evaluations with different LLM backbones, conduct ablation studies, analyze consistency with human judgments, and investigate performance across difficulty and attack-stage levels.

    \item \textbf{Appendix~\ref{app:experimental_details}: Experimental Details.}
    We provide detailed descriptions of baseline configurations, validation environments, and experimental settings.

    \item \textbf{Appendix~\ref{app:Subjective Metrics}: Details of Semantic Evaluation via LLM-as-a-Judge.}
    We describe the scoring criteria, multi-dimensional evaluation protocol, and prompts used for semantic assessment.

    \item \textbf{Appendix~\ref{app:att_be_ta_bo}: Attack Behavior Taxonomy and Boundaries.}
    We present the complete attack behavior taxonomy, discuss validation boundaries, and summarize future extensions of CyberClear.

    \item \textbf{Appendix~\ref{app:case}: More Cases.}
    We provide additional qualitative examples of attack-chain provenance reconstruction and error analysis.

\end{itemize}

\section{Additional Experiments}
\label{app:additional_results}

\subsection{Evaluation with Additional LLM Backbones}
\label{app:other_llm}

\begin{table*}[h]
\setlength{\belowcaptionskip}{5pt}
\caption{Additional results of representative agent frameworks on CyberClear with different LLM backbones.}
\centering
\small
\setlength{\tabcolsep}{1.2pt}
\renewcommand{\arraystretch}{1.2}
\resizebox{\textwidth}{!}{
\begin{tabular}{lc|cccc|ccccc}
\toprule

\multirow{2}{*}{\textbf{Methods}}
& \multirow{2}{*}{}
& \multicolumn{4}{c|}{\textbf{Code-Level Metrics}}
& \multicolumn{5}{c}{\textbf{Semantic Metrics}}
\\

\cmidrule(lr){3-6}
\cmidrule(lr){7-11}

& 
& BLEU 
& ROUGE-L 
& CodeBLEU 
& Pass@1
& Single 
& Loose 
& Strict 
& Detail 
& Scenario
\\

\midrule

\multirow{2}{*}{OpenClaw (Single)}
& \includegraphics[height=0.33cm]{img/gpt.pdf}
& 0.3617 & 0.5236 & 0.3019 & 0.8113
& 0.4509 & 0.3588 & 0.2253 & 0.2602 & 0.3209 \\

& \includegraphics[height=0.33cm]{img/gemini.pdf}
& 0.3428 & 0.4975 & 0.3659 & 0.8267
& 0.6521 & 0.5748 & 0.4146 & 0.3917 & 0.4615 \\

\midrule

\multirow{2}{*}{OpenClaw (Multiple)}
& \includegraphics[height=0.33cm]{img/gpt.pdf}
& 0.4483 & 0.6027 & 0.4872 & 0.9085
& 0.5694 & 0.4592 & 0.2982 & 0.3279 & 0.4293 \\

& \includegraphics[height=0.33cm]{img/gemini.pdf}
& 0.4237 & 0.5718 & 0.5526 & 0.9178
& 0.5894 & 0.5127 & 0.3812 & 0.3526 & 0.4098 \\

\midrule

\multirow{2}{*}{AgentVerse}
& \includegraphics[height=0.33cm]{img/gpt.pdf}
& 0.3964 & 0.5819 & 0.5298 & 0.9289
& 0.6383 & 0.5314 & 0.3510 & 0.3650 & 0.4869 \\

& \includegraphics[height=0.33cm]{img/gemini.pdf}
& 0.3749 & 0.5521 & 0.5963 & 0.9378
& 0.5486 & 0.4725 & 0.3497 & 0.3264 & 0.3618 \\

\midrule

\multirow{2}{*}{LLM-Debate}
& \includegraphics[height=0.33cm]{img/gpt.pdf}
& 0.4389 & 0.5651 & 0.5726 & 0.8533
& 0.7407 & 0.6388 & 0.4284 & 0.4178 & 0.5727 \\

& \includegraphics[height=0.33cm]{img/gemini.pdf}
& 0.4146 & 0.5387 & 0.6385 & 0.8711
& 0.6418 & 0.5589 & 0.4236 & 0.3895 & 0.4587 \\

\midrule

\multirow{2}{*}{MultiPersona}
& \includegraphics[height=0.33cm]{img/gpt.pdf}
& 0.4075 & 0.5943 & 0.3441 & 0.9178
& 0.4299 & 0.3581 & 0.2263 & 0.2476 & 0.2945 \\

& \includegraphics[height=0.33cm]{img/gemini.pdf}
& 0.3893 & 0.5685 & 0.4148 & 0.9267
& 0.6635 & 0.5864 & 0.4281 & 0.3972 & 0.4716 \\

\midrule

\multirow{2}{*}{LLM-Blender}
& \includegraphics[height=0.33cm]{img/gpt.pdf}
& 0.4496 & 0.5868 & 0.4985 & 0.8622
& 0.5911 & 0.4816 & 0.3147 & 0.3397 & 0.4475 \\

& \includegraphics[height=0.33cm]{img/gemini.pdf}
& 0.4261 & 0.5584 & 0.5689 & 0.8800 & 0.5962
& 0.5187 & 0.3864 & 0.3615 & 0.4219 \\

\midrule

\multirow{2}{*}{DyLAN}
& \includegraphics[height=0.33cm]{img/gpt.pdf}
& 0.4362 & 0.5762 & 0.6144 & 0.8622
& 0.7863 & 0.6842 & 0.4583 & 0.4366 & 0.6127 \\

& \includegraphics[height=0.33cm]{img/gemini.pdf}
& 0.4584 & 0.5783 & 0.6643 & 0.8844
& 0.7583 & 0.6273 & 0.4258 & 0.4538 & 0.6029 \\

\midrule

\multirow{2}{*}{MacNet}
& \includegraphics[height=0.33cm]{img/gpt.pdf}
& 0.3908 & 0.5559 & 0.4066 & 0.8467
& 0.4413 & 0.2981 & 0.1763 & 0.2564 & 0.3343 \\

& \includegraphics[height=0.33cm]{img/gemini.pdf}
& 0.3933 & 0.5578 & 0.5444 & 0.8466
& 0.5742 & 0.4674 & 0.3567 & 0.4523 & 0.4196 \\

\midrule

\multirow{2}{*}{CAMEL}
& \includegraphics[height=0.33cm]{img/gpt.pdf}
& 0.3902 & 0.5873 & 0.3384 & 0.9222
& 0.4264 & 0.3596 & 0.2278 & 0.2454 & 0.2894 \\

& \includegraphics[height=0.33cm]{img/gemini.pdf}
& 0.3887 & 0.5842 & 0.4219 & 0.9222
& 0.4753 & 0.3606 & 0.2563 & 0.2505 & 0.3009 \\

\midrule

\multirow{2}{*}{AgentScope-V2}
& \includegraphics[height=0.33cm]{img/gpt.pdf}
& 0.4642 & 0.6073 & 0.6513 & 0.9867
& 0.6646 & 0.5702 & 0.3934 & 0.4035 & 0.4998 \\

& \includegraphics[height=0.33cm]{img/gemini.pdf}
& 0.5035 & 0.6435 & 0.7311 & 0.8622
& 0.7095 & 0.6889 & 0.4785 & 0.5557 & 0.6278 \\

\midrule

\multirow{2}{*}{CyberProvenance (Ours)}
& \includegraphics[height=0.33cm]{img/gpt.pdf}
& 0.4871 & 0.6246 & 0.6673 & 0.9822
& 0.6748 & 0.5792 & 0.3974 & 0.4046 & 0.5098 \\

& \includegraphics[height=0.33cm]{img/gemini.pdf}
& 0.4896 & 0.6738 & 0.6941 & 0.9911 
& 0.8485 & 0.7186 & 0.6179 & 0.5843 & 0.6897 \\

\bottomrule

\end{tabular}
}

\label{tab:other_llm}
\vspace{8pt}
\end{table*}

Table~\ref{tab:other_llm} reports additional results with GPT-4o-mini and Gemini-3.5-Flash-Lite to examine the robustness of different agent frameworks across additional LLM backbones. 
The results show substantial performance variation across backbones, indicating that provenance is sensitive to the underlying model capability. 
With Gemini-3.5-Flash-Lite, CyberProvenance achieves the strongest performance across all five Semantic Metrics metrics while maintaining competitive code-level similarity. 
In contrast, performance with GPT-4o-mini is generally lower, particularly on Strict, Detail, and Scenario, highlighting the difficulty of reconstructing complete attack chains with weaker backbones. 
These results further show that both the agent framework and the underlying LLM contribute substantially to reliable APT attack chain provenance.

\subsection{Ablation Studies}

\begin{table*}[h]
\setlength{\belowcaptionskip}{5pt}
\caption{
Ablation study of CyberProvenance on CyberClear using GPT-5.6-Terra.
Red values indicate performance drops relative to the full model.
}
\centering
\small
\setlength{\tabcolsep}{2.2pt}
\renewcommand{\arraystretch}{1.25}

\resizebox{\textwidth}{!}{
\begin{tabular}{l|cccc|ccccc}
\toprule

\multirow{2}{*}{\textbf{Variants}}
& \multicolumn{4}{c|}{\textbf{Code-Level Metrics}}
& \multicolumn{5}{c}{\textbf{Semantic Metrics}}
\\

\cmidrule(lr){2-5}
\cmidrule(lr){6-10}

& BLEU
& ROUGE-L
& CodeBLEU
& Pass@1
& Single
& Loose
& Strict
& Detail
& Scenario
\\

\midrule

\rowcolor{blue!7}
\textbf{CyberProvenance}
& \textbf{0.4952}
& \textbf{0.6968}
& \textbf{0.7016}
& \textbf{0.9333}
& \textbf{0.8897}
& \textbf{0.7438}
& \textbf{0.6337}
& \textbf{0.6124}
& \textbf{0.7263}
\\

\midrule

w/o Evidence Memory
& 0.4510 {\scriptsize\textcolor{red!70!black}{$\downarrow$0.0442}}
& 0.6423 {\scriptsize\textcolor{red!70!black}{$\downarrow$0.0545}}
& 0.6554 {\scriptsize\textcolor{red!70!black}{$\downarrow$0.0462}}
& 0.9044 {\scriptsize\textcolor{red!70!black}{$\downarrow$0.0289}}
& 0.8421 {\scriptsize\textcolor{red!70!black}{$\downarrow$0.0476}}
& 0.6845 {\scriptsize\textcolor{red!70!black}{$\downarrow$0.0593}}
& 0.5578 {\scriptsize\textcolor{red!70!black}{$\downarrow$0.0759}}
& 0.5236 {\scriptsize\textcolor{red!70!black}{$\downarrow$0.0888}}
& 0.6502 {\scriptsize\textcolor{red!70!black}{$\downarrow$0.0761}}
\\

w/o Execution Validation
& 0.4598 {\scriptsize\textcolor{red!70!black}{$\downarrow$0.0354}}
& 0.6492 {\scriptsize\textcolor{red!70!black}{$\downarrow$0.0476}}
& 0.6615 {\scriptsize\textcolor{red!70!black}{$\downarrow$0.0401}}
& 0.8978 {\scriptsize\textcolor{red!70!black}{$\downarrow$0.0355}}
& 0.8470 {\scriptsize\textcolor{red!70!black}{$\downarrow$0.0427}}
& 0.6780 {\scriptsize\textcolor{red!70!black}{$\downarrow$0.0658}}
& 0.5485 {\scriptsize\textcolor{red!70!black}{$\downarrow$0.0852}}
& 0.5200 {\scriptsize\textcolor{red!70!black}{$\downarrow$0.0924}}
& 0.6425 {\scriptsize\textcolor{red!70!black}{$\downarrow$0.0838}}
\\

w/o Feedback Refinement
& 0.4825 {\scriptsize\textcolor{red!70!black}{$\downarrow$0.0127}}
& 0.6811 {\scriptsize\textcolor{red!70!black}{$\downarrow$0.0157}}
& 0.6889 {\scriptsize\textcolor{red!70!black}{$\downarrow$0.0127}}
& 0.9178 {\scriptsize\textcolor{red!70!black}{$\downarrow$0.0155}}
& 0.8734 {\scriptsize\textcolor{red!70!black}{$\downarrow$0.0163}}
& 0.7169 {\scriptsize\textcolor{red!70!black}{$\downarrow$0.0269}}
& 0.5983 {\scriptsize\textcolor{red!70!black}{$\downarrow$0.0354}}
& 0.5716 {\scriptsize\textcolor{red!70!black}{$\downarrow$0.0408}}
& 0.6885 {\scriptsize\textcolor{red!70!black}{$\downarrow$0.0378}}
\\

\bottomrule
\end{tabular}
}

\label{tab:ablation}
\vspace{8pt}
\end{table*}

To quantify the contribution of each harness component, we remove Evidence Memory, Execution Validation, and Feedback Refinement individually while keeping all other settings unchanged. 
As shown in Table~\ref{tab:ablation}, removing Execution Validation causes the largest degradation, reducing the average Semantic Metrics score from 0.7212 to 0.6472, with especially large drops in Strict, Detail, and Scenario.
Removing Evidence Memory yields a comparable decline to 0.6516 and notably reduces Strict and Detail by 0.0759 and 0.0888, respectively, showing the importance of maintaining consistent evidence across long-context logs. 
Removing Feedback Refinement leads to a smaller but consistent decrease, with the average semantic score falling to 0.6897 and all evaluation metrics degrading relative to the full model. 
These results demonstrate that Evidence Memory and Execution Validation provide the main gains, while Feedback Refinement further improves reconstruction quality through iterative correction.

\subsection{Consistency with Human Judgments}

\begin{figure*}[h]
    \centering
    \includegraphics[width=\linewidth]{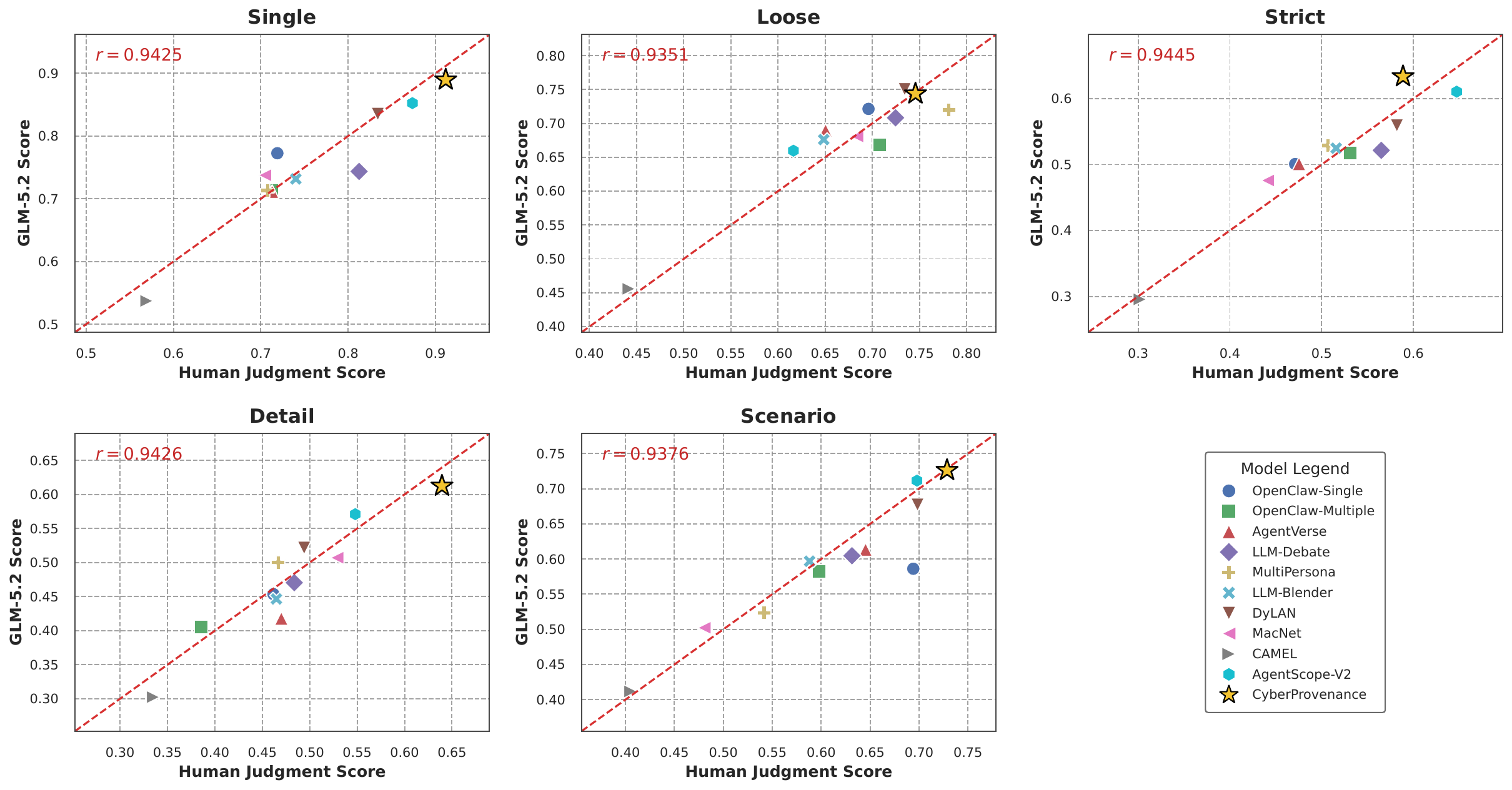}
    \caption{
Consistency between GPT-5.6-Terra-based evaluation and human judgments.
Two human experts independently evaluate generated provenance graphs, with each evaluation repeated three times and averaged.
}
    \label{fig:human_consistency}
    \vspace{-3mm}
\end{figure*}

To evaluate the reliability of our semantic evaluation method, we compare the model-based scores with human judgments.
Two human experts independently assess the generated provenance graphs, and each expert repeats the evaluation three times; the final human score is obtained by averaging all judgments.
We use the GPT-5.6-Terra evaluation results for this analysis and measure the correlation between automatic scores and human assessments.
As shown in Figure~\ref{fig:human_consistency}, the scores produced by our evaluator exhibit strong consistency with human judgments across all five dimensions.
The Pearson correlation coefficients reach 0.9425, 0.9351, 0.9445, 0.9426, and 0.9376 for Single, Loose, Strict, Detail, and Scenario, respectively.
These results indicate that our evaluation method can reliably approximate human assessments while providing a scalable alternative for evaluating attack-chain provenance reconstruction.

\subsection{Additional Performance Analysis Across Difficulty Levels}

\begin{figure*}[t]
    \centering
    \includegraphics[width=\linewidth]{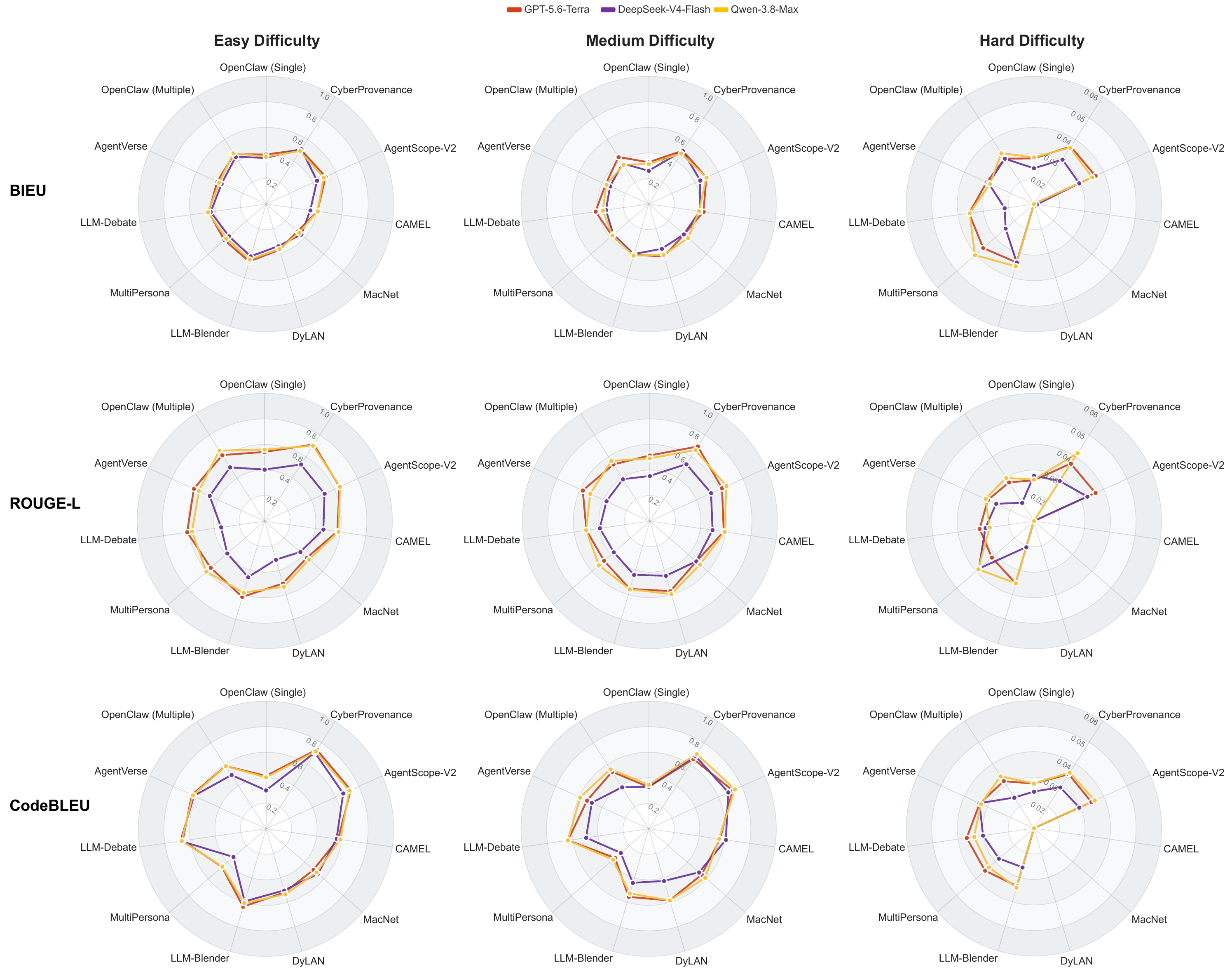}
    \caption{Code-Level similarity across difficulty levels using BLEU, ROUGE-L, and CodeBLEU. Columns correspond to easy, medium, and hard instances, and each radar chart compares agent frameworks under three LLM backbones.}
    \label{fig:difficulty_code_similarity}
\end{figure*}

The additional results examine how code-level similarity varies with task difficulty across different agent frameworks and LLM backbones. 
BLEU, ROUGE-L, and CodeBLEU remain relatively stable for most methods on easy and medium instances, but decrease substantially on hard instances. 
As shown in Figure~\ref{fig:difficulty_code_similarity}, the decline becomes more pronounced as the complexity of attack-chain reconstruction increases. 
Hard instances also produce larger performance gaps among different agent frameworks, indicating greater differences in their ability to preserve the structure of the target provenance graphs. 
This trend is consistently observed across the three LLM backbones, although the extent of degradation varies across models. 
These results further show that increasing task difficulty challenges both the semantic reconstruction of attack chains and the structural consistency of the generated graphs.


\subsection{Additional Performance Analysis Across Attack Stage Levels}

\begin{figure*}[t]
    \centering
    \includegraphics[width=\linewidth]{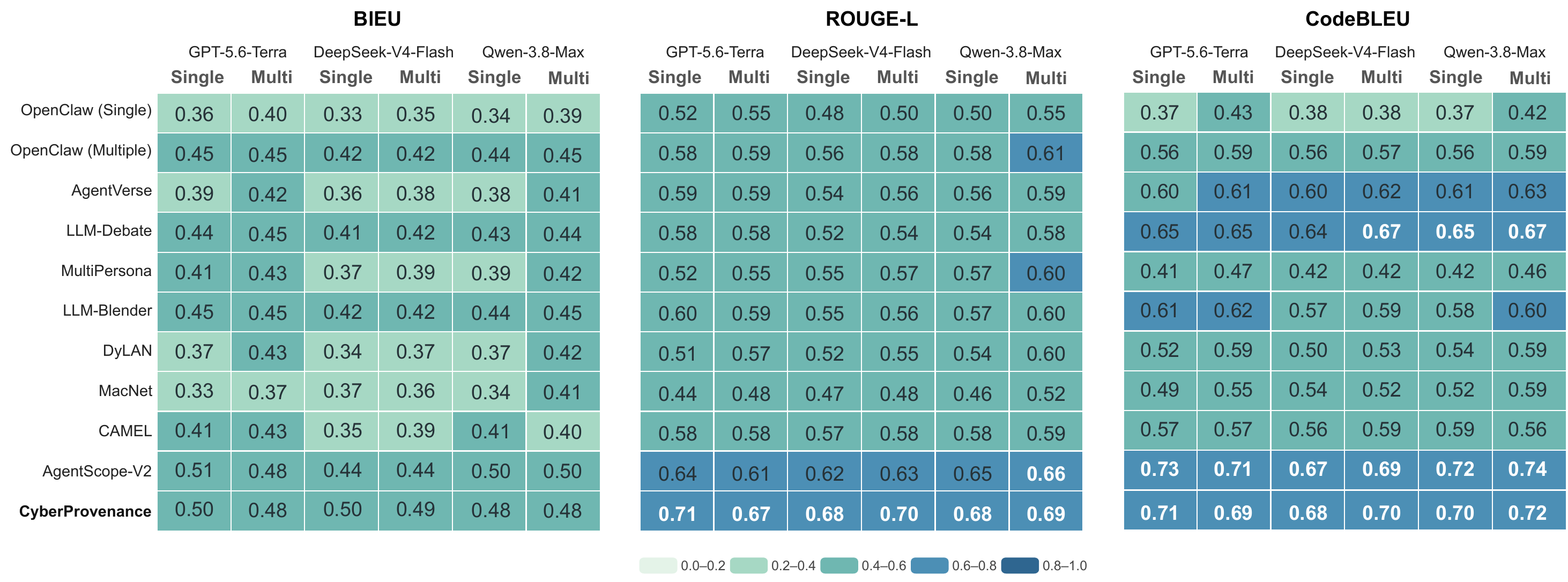}
    \caption{Code-Level similarity across single-stage and multi-stage instances using BLEU, ROUGE-L, and CodeBLEU. Columns correspond to three LLM backbones, and each heatmap compares agent frameworks under single-stage and multi-stage settings.}
    \label{fig:stage_code_similarity}
    \vspace{-3mm}
\end{figure*}

Across the three code-level metrics, agent frameworks exhibit clear differences between single-stage and multi-stage attack-chain reconstruction. 
Most methods show relatively stable BLEU and ROUGE-L scores across the two settings, while larger variations can be observed in CodeBLEU. 
As shown in Figure~\ref{fig:stage_code_similarity}, AgentScope-V2 and CyberProvenance maintain consistently high scores across different LLM backbones, indicating stronger structural fidelity in the generated provenance graphs. 
In contrast, several general-purpose agent frameworks achieve noticeably lower similarity, especially when reconstructing more complex multi-stage attack chains. 
The comparison across GPT-5.6-Terra, DeepSeek-V4-Flash, and Qwen-3.8-Max further shows that the same agent framework can exhibit different levels of robustness depending on the underlying LLM. 
These results suggest that reliable provenance graph generation depends not only on the backbone model, but also on whether the agent framework provides effective support for iterative refinement, evidence consistency, and cybersecurity-oriented validation.

\section{Experimental Details}
\label{app:experimental_details}

\subsection{Baselines}
\label{app:baselines}

For all baselines, we set the temperature to 0.2 and use top-$p=1.0$. The remaining configuration details for each baseline are provided below:

\textbf{OpenClaw (Single)~\citep{openclaw2026}.}
We configure OpenClaw with a single LLM agent for security log analysis, attack chain reconstruction, and DOT generation, using Web Search when needed and allowing up to three rounds of self verification and refinement.
We utilize the official implementation available at \url{https://github.com/openclaw/openclaw}.

\textbf{OpenClaw (Multiple).}
We build three collaborative OpenClaw agents for attack evidence extraction, DOT code generation, and output verification, respectively. Web Search is restricted to ATT\&CK identifier and name verification, and the system performs up to three rounds of generation, review, and refinement.
We use the official code released at \url{https://github.com/openclaw/openclaw}.

\textbf{AgentVerse~\citep{chen2024agentverse}.}
We configure AgentVerse with four expert agents that collaborate under a predefined protocol for log analysis and attack chain reconstruction. 
The agents jointly analyze evidence, propose candidate reconstructions, and consolidate the final DOT graph.
The source code is available at \url{https://github.com/OpenBMB/AgentVerse/}.

\textbf{LLM-Debate~\citep{du2023improving}.}
We configure LLM-Debate with three agents and two debate rounds. Each agent independently generates an attack-chain candidate, then revises it after reviewing the other agents’ outputs, with the final results consolidated into a single reconstruction. We use the official implementation available at \url{https://github.com/composable-models/llm_multiagent_debate}.

\textbf{MultiPersona~\citep{wang2024unleashing}.}
In our experiments, MultiPersona uses a single LLM instance to simulate three task-specific personas for attack-chain reconstruction. The personas interact over multiple rounds to analyze attack evidence and collaboratively refine the reconstructed chain. The experiments are conducted using the implementation provided at \url{https://github.com/MikeWangWZHL/Solo-Performance-Prompting}.

\textbf{LLM-Blender~\citep{jiang2023llm}.}
For LLM-Blender, three generator LLMs produce 11 candidate reconstructions, which are pairwise ranked by PairRanker. The top three candidates ($K=3$) are retained and fused by GenFuser to produce the final DOT graph. Our experiments build on the implementation released at \url{https://github.com/yuchenlin/LLM-Blender}.

\textbf{DyLAN~\citep{liu2024dylan}.}
We enable DyLAN's dynamic agent selection, multi-round interaction, and early stopping for attack-chain reconstruction. The corresponding implementation is obtained from \url{https://github.com/SALT-NLP/DyLAN}.

\textbf{MacNet~\citep{qian2025scaling}.}
In the MacNet setting, agents are organized as a directed acyclic collaboration network and interact following the topological order to derive the final reconstruction. We conduct the experiments with the MacNet implementation from \url{https://github.com/OpenBMB/ChatDev/tree/macnet}.

\textbf{CAMEL~\citep{li2023camel}.}
For CAMEL, we assign predefined roles to the participating agents and allow them to collaboratively reconstruct the attack chain through autonomous role-playing dialogue. The experimental setup is implemented with the codebase at \url{https://github.com/camel-ai/camel}.

\textbf{AgentScope-V2~\citep{agentscope,agentscope_v1}.}
AgentScope-V2 is configured with role-specialized agents that exchange intermediate evidence and reconstruction results through structured messages, allowing later agents to refine earlier outputs before producing the final graph. We implement this baseline using the AgentScope repository at \url{https://github.com/agentscope-ai/agentscope}.

\subsection{Attack Validation Environment}
\label{app:tool}

For environment-based validation, we select execution procedures using the predicted ATT\&CK Technique identifiers, while obtaining platform and runtime settings from predefined environment configurations. The configurations of the execution and telemetry tools are provided below:

\textbf{Atomic Red Team~\citep{atomicredteam}.}
We use Atomic Red Team to reproduce individual attack steps.
For each predicted Technique identifier, we retrieve candidate tests from a predefined catalog and select compatible implementations according to their supported platforms, execution requirements, and dependencies.
Execution results and the corresponding telemetry are retained for subsequent comparison with the observed evidence.
We use the official implementation available at \url{https://github.com/redcanaryco/atomic-red-team}.

\textbf{CALDERA~\citep{apachecaldera}.}
We use CALDERA to orchestrate multi-step attack chains with execution dependencies.
Predicted Technique identifiers are mapped to predefined abilities, whose requirements and parsers specify execution prerequisites and extract facts from procedure outputs.
The execution agents and planner are specified by the experiment configuration.
Operation records and execution traces are retained to assess step completion and examine dependencies between steps.
We use the official implementation available at \url{https://github.com/apache/caldera}.

\textbf{Ludus~\citep{ludus}.}
We use Ludus with Proxmox to provision isolated virtual machines from predefined environment templates.
The range configuration specifies the operating systems, network topology, services, and telemetry components required by the selected procedures.
Testing mode is configured to block communication outside the experimental range, and snapshots provide reproducible starting states for validation runs.
The official documentation is available at \url{https://docs.ludus.cloud/}.

\textbf{Sysmon~\citep{sysmon}.}
We deploy Sysmon on virtual machines to collect host activity under a predefined event-filtering configuration.
The collected events cover process creation, network connections, and file activity, providing host-level observations for comparison with the reconstructed attack steps.
We use the official distribution available at \url{https://learn.microsoft.com/en-us/sysinternals/downloads/sysmon}.

\textbf{Linux Audit~\citep{linuxaudit}.}
We deploy Linux Audit on virtual machines and configure audit rules for the system calls and file operations relevant to the selected procedures.
The audit daemon collects the resulting records, including timestamps, process identifiers, user identifiers, and operation outcomes where available.
These records provide execution evidence for validating attack steps on Linux hosts.
We use the official implementation available at \url{https://github.com/linux-audit/audit-userspace}.

\textbf{Zeek~\citep{zeek}.}
We use Zeek to monitor traffic visible on the experimental range's designated capture interface.
Connection records and protocol-specific logs, including DNS and HTTP records where applicable, complement the host telemetry collected by Sysmon and Linux Audit.
The resulting network observations are associated with validation runs and incorporated into the structured validation reports.
We use the official implementation available at \url{https://github.com/zeek/zeek}.

\section{Details of Semantic Evaluation via LLM-as-a-Judge}
\label{app:Subjective Metrics}

\subsection{Scoring Criteria}
\label{app:scoring criteria}

We evaluate each reference attack step by aligning it with the predicted content and assessing the degree of semantic agreement. 
Based on this comparison, each step is categorized as \textbf{Correct}, \textbf{Minor Partial}, \textbf{Moderate Partial}, \textbf{Major Partial}, \textbf{Incorrect}, or \textbf{Missing}. We then assign an integer raw score in $[0,100]$ according to the step-level scoring criteria below and normalize it to $[0,1]$ by dividing by 100.
\begin{itemize}[leftmargin=*, itemsep=3pt, topsep=0pt]
    \item \textbf{(80,100] (Correct):}
    The reference attack event is accurately reconstructed. Its actor, core
    action, target, and overall attack meaning are preserved. A slightly lower
    score within this range may be assigned when only limited non-critical
    details are omitted or harmlessly imprecise.

    \item \textbf{(60,80] (Minor Partial):}
    The same event and its core behavior are clearly identifiable, but a
    secondary behavioral qualifier, execution method, local outcome, or causal
    explanation is incomplete. These differences do not alter the identity,
    attack stage, or principal meaning of the event.

    \item \textbf{(40,60] (Moderate Partial):}
    The same event remains identifiable, but at least one core semantic
    component of the actor, action, target, or outcome is missing, incorrect,
    or ambiguous. Consequently, the attack process is incomplete, although it
    has not changed into a different event.

    \item \textbf{(20,40] (Major Partial):}
    The prediction is substantially incomplete and preserves only a broad core
    behavior, with most important information missing or inaccurate.

    \item \textbf{(0,20] (Incorrect):}
    The predicted event substantively conflicts with the reference event or
    exhibits only a very weak semantic correspondence.

    \item \textbf{0 (Missing):}
    The prediction contains no behavior with a meaningful semantic
    correspondence to the reference attack event.
\end{itemize}

\subsection{Multi-dimensional Semantic Assessment}

In this section, we provide detailed descriptions of our multi-dimensional semantic evaluation method.
Given a reference provenance graph $G$ and a predicted provenance graph $\hat{G}$, the judge directly compares their DOT code, treating the reference graph as authoritative.
Differences in node identifiers, DOT statement order, graph layout, colors, shapes, and other purely visual properties are ignored, while entities and attributes, such as commands, file paths, IP addresses, timestamps, PIDs, services, and ATT\&CK identifiers, are evaluated when they carry forensic significance.
We use GLM-5.2 as the judge model with the temperature set to 0, and require each judgment to provide supporting evidence from both the reference and predicted graphs whenever available.

\paragraph{Single-Step Judge.}
The Single-Step Judge evaluates the semantic agreement between the reference event and the predicted content for single-step samples.
It directly assigns a real-valued score in the range $[0,1]$ based on whether the predicted event preserves the core action, target, and overall meaning of the reference event.

\paragraph{Loose Judge.}
The Loose Judge evaluates multi-step samples by independently measuring the semantic coverage of each reference attack step without considering temporal order.
The sample-level score is computed as the mean of all step-level scores.
A predicted step may cover multiple reference steps only when it contains distinct sub-actions corresponding to these steps, and the same predicted sub-action cannot be reused to match multiple reference steps.
Unsupported predicted actions cannot match reference steps and do not contribute to the score.

\paragraph{Strict Judge.}
The Strict Judge evaluates whether the predicted attack chain preserves the temporal progression of the reference chain.
Let $K$ denote the number of reference steps in the longest consecutively correct prefix, and let $N$ denote the total number of reference steps.
The prefix is counted in reference-step units rather than predicted-step positions, allowing one predicted step to cover multiple consecutive reference steps when its distinct sub-actions appear in the correct semantic order.

The prefix terminates at the first partial or incorrect reference event, or when an unsupported attack action is inserted before the reference chain is completed.
If all reference steps are correctly covered, $U$ denotes the number of unsupported attack actions appended after the completed chain.
The Strict Judge score is defined as:

\begin{equation}
S_{\mathrm{strict}} =
\begin{cases}
\dfrac{K}{N}, & K<N,\\[2mm]
\dfrac{2N}{2N+U}, & K=N.
\end{cases}
\end{equation}

A completely correct chain without unsupported appended actions receives a score of 1.
Unsupported entities, relations, and attributes are evaluated separately by the Detail Judge.

\paragraph{Detail Judge.}
The Detail Judge evaluates the fidelity of graph-level details, including entities, directed relations, and semantic attributes.
Purely visual properties, such as node colors, shapes, and layout information, are ignored.
Reference and predicted elements are aligned one-to-one within each category.

A fully correct match receives a weight of $1.0$, while partial matches receive weights of $0.75$, $0.50$, or $0.25$ according to the preserved semantic information.
Incorrect matches, uncovered reference elements, and unsupported predicted elements receive zero weight.

For each category $c\in\{\mathrm{ent},\mathrm{rel},\mathrm{attr}\}$, we compute:

\begin{equation}
F_c =
\frac{2TP_c^{w}}
{N_c^{\mathrm{ref}}+N_c^{\mathrm{pred}}},
\qquad
S_{\mathrm{detail}} =
\sum_c w_cF_c ,
\end{equation}

where $TP_c^{w}$ denotes the weighted semantic matches, and $N_c^{\mathrm{ref}}$ and $N_c^{\mathrm{pred}}$ denote the numbers of reference and predicted elements in category $c$.
We set $w_{\mathrm{ent}}=0.35$, $w_{\mathrm{rel}}=0.45$, and $w_{\mathrm{attr}}=0.20$.
The final Detail-Judge score is normalized to $[0,1]$.

\paragraph{Scenario Judge.}
The Scenario Judge evaluates the overall consistency between the reference and predicted attack descriptions extracted from DOT files.
It assigns a real-valued score in the range $[0,1]$ according to whether the two descriptions convey the same attack process, target, method, progression, and outcome.
The judge allows synonymous expressions, semantic paraphrases, and harmless generalizations.
However, contradictions, unsupported core attack stages, and fabricated attack objectives reduce the final score.

\subsection{LLM-as-a-Judge Prompts}
\label{app:llm_judge_prompts}



\tcbinputlisting{
  promptbox,
  title={System Prompt},
  title after break={System Prompt (continued)},
  listing file={./prompts/system_prompt.txt}
}
\vspace{5pt}
\promptcaption{System prompt defining the evaluator's role, evidence constraints, and JSON-only response requirements.}{fig:judge_system_prompt}


\tcbinputlisting{
  promptbox,
  title={General Evaluation Protocol},
  title after break={General Evaluation Protocol (continued)},
  listing file={./prompts/general_protocol.txt}
}
\vspace{5pt}
\promptcaption{General evaluation protocol specifying shared principles for evidence-grounded semantic comparison of reference and predicted provenance graphs.}{fig:judge_general_protocol}

\tcbinputlisting{
  promptbox,
  title={Single-Step and Loose Judge Prompt},
  title after break={Single-Step and Loose Judge Prompt (continued)},
  listing file={./prompts/step_judge_prompt.txt}
}
\vspace{5pt}
\promptcaption{Single-Step and Loose Judge prompt for evaluating individual attack steps and unordered semantic coverage of multi-step attack chains.}{fig:judge_step_prompt}

\tcbinputlisting{
  promptbox,
  title={Strict Judge Prompt},
  title after break={Strict Judge Prompt (continued)},
  listing file={./prompts/strict_judge_prompt.txt}
}
\vspace{5pt}
\promptcaption{Strict Judge prompt for evaluating the consecutive correct prefix of an attack chain under ordered sub-action alignment.}{fig:judge_strict_prompt}

\tcbinputlisting{
  promptbox,
  title={Detail Judge Prompt},
  title after break={Detail Judge Prompt (continued)},
  listing file={./prompts/detail_judge_prompt.txt}
}
\vspace{5pt}
\promptcaption{Detail Judge prompt for evaluating entity, relation, and attribute coverage through one-to-one reference--prediction matching.}{fig:judge_detail_prompt}

\tcbinputlisting{
  promptbox,
  title={Scenario Judge Prompt},
  title after break={Scenario Judge Prompt (continued)},
  listing file={./prompts/scenario_judge_prompt.txt}
}
\vspace{5pt}
\promptcaption{Scenario Judge prompt for holistic evaluation of semantic agreement between reference and predicted attack scenarios.}{fig:judge_scenario_prompt}

\tcbinputlisting{
  promptbox,
  title={Required JSON Output Schema},
  title after break={Required JSON Output Schema (continued)},
  listing file={./prompts/output_schema.txt}
}
\vspace{5pt}
\promptcaption{Required JSON output schema for reporting applicable metric scores, score calculations, and supporting evaluation details.}{fig:judge_output_schema}

\section{Attack Behavior Taxonomy and Boundaries}
\label{app:att_be_ta_bo}
\subsection{Complete List of Attack Behaviors}
\label{app:behaviors}

Table~\ref{tab:attack_behaviors} presents the complete attack behaviors in CyberClear. 
We normalize the annotated attack behaviors according to their MITRE ATT\&CK techniques and merge instances with identical technique sets. 
This process yields 89 distinct attack behaviors from 450 instances, including 46 single-technique behaviors and 43 composite behaviors. 
Composite behaviors capture attacks involving multiple ATT\&CK techniques and thus reflect more complex attacker activity patterns.

We further evaluate the attack-behavior coverage of different agent frameworks on CyberClear. 
For each behavior, we compute the compilation success rate over all associated instances and regard a generated provenance graph as successful when its DOT code can be compiled correctly. 
As shown in Table~\ref{tab:gpt_behavior_coverage}, a behavior is considered supported when at least 80\% of its instances are successfully compiled. 
This analysis provides a finer-grained view of model capability by showing which attack behaviors can be handled reliably beyond the aggregate Pass@1 score.

\subsection{Method Boundaries}

We ground the concrete validation implementation in the observed log evidence, while using ATT\&CK Techniques to specify the corresponding attack behavior categories. 
Because the same Technique may admit multiple implementations across operating systems, tools, and versions, the selected implementation must be compatible with the execution context and evidence reflected in the logs. 
We obtain infrastructure-dependent runtime parameters from predefined configurations and validate inter-step dependencies exclusively using facts and evidence generated during the current execution. 
After each local refinement, we replay the attack chain from the initial snapshot to prevent previously generated facts from affecting subsequent dependency verification. 
We conduct all validation in authorized private environments with external network access blocked, predefined target scopes, and non-destructive execution constraints.

\subsection{Future Work}

Future work will explore adaptive cyber harnesses that dynamically optimize evidence retrieval, validation strategies, tool selection, and agent collaboration based on accumulated experience.
Meanwhile, we plan to extend the scale and scope of CyberClear by incorporating more APT campaigns, ATT\&CK behaviors, and longer multi-stage attack chains.
We will broaden the benchmark to cover more diverse environments, including host, network, cloud, and container settings.
We also aim to include more challenging observations with noisy logs, missing evidence, and cross-source inconsistencies.
In addition, we will enrich the benchmark with finer-grained annotations for attack-step dependencies and evidence alignment.
Finally, we will evaluate emerging agent frameworks and foundation models to provide a more comprehensive understanding of their capabilities in complete attack-chain reconstruction.

\section{More Cases}
\label{app:case}

This section presents additional cases from CyberClear.

\vspace{8pt}
\begin{figure*}[h]
    \centering
    \includegraphics[width=0.88\linewidth]{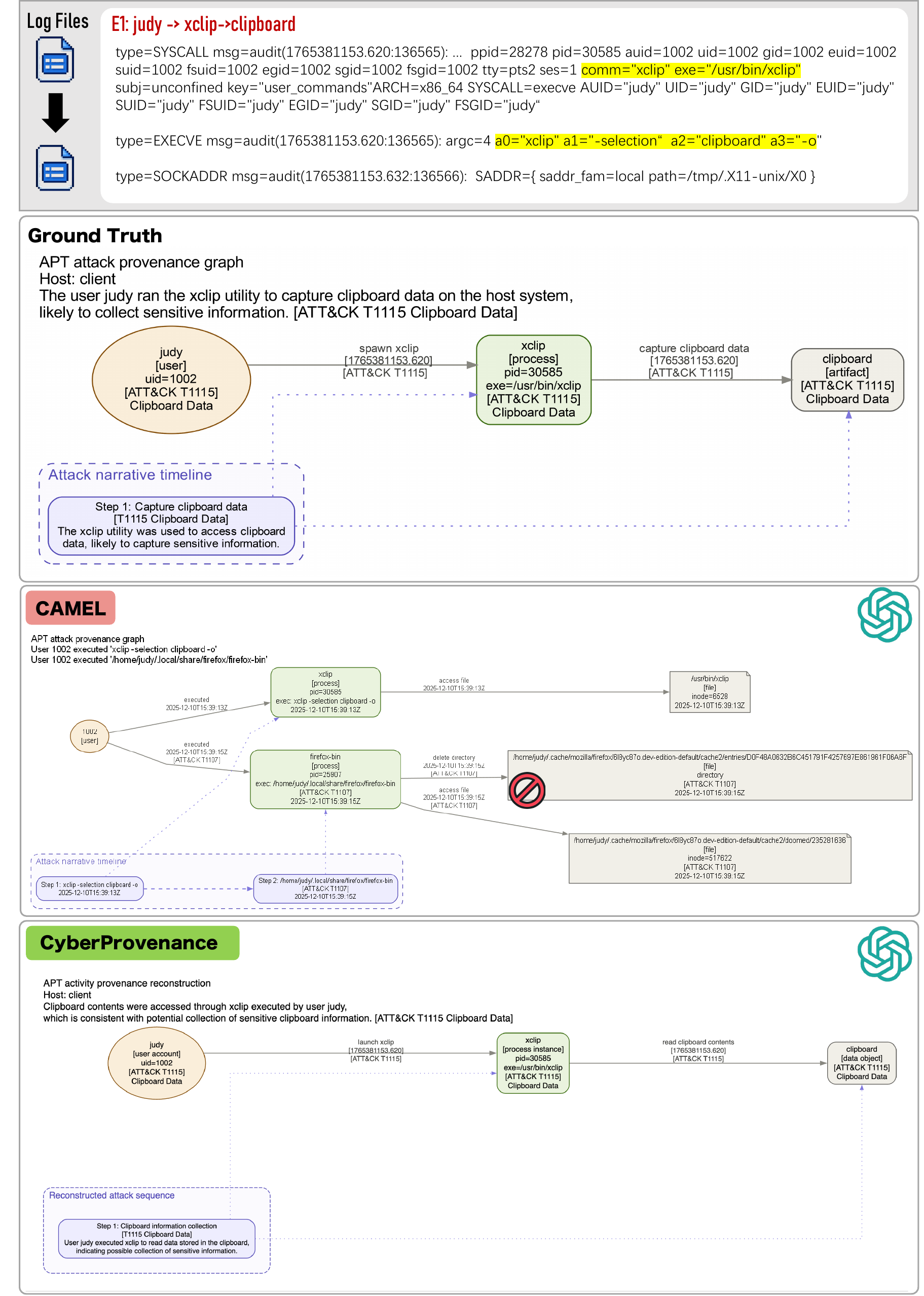}
    \vspace{-6pt}
    \caption{
Case study of attack-chain provenance.
LLM-Debate incorrectly omits the \texttt{bash} $\rightarrow$ \texttt{stty} dependency and its terminal-configuration step, despite explicit audit evidence highlighted in yellow: \texttt{comm="stty"}, \texttt{exe="/usr/bin/stty"}, and \texttt{stty rows 38 columns 116}.
CyberProvenance correctly aligns this evidence with the corresponding entity relation and attack step, producing a provenance graph consistent with the observed execution.
}
    \label{fig:case21}
\end{figure*}

\begin{figure*}[h]
    \centering
    \includegraphics[width=\linewidth]{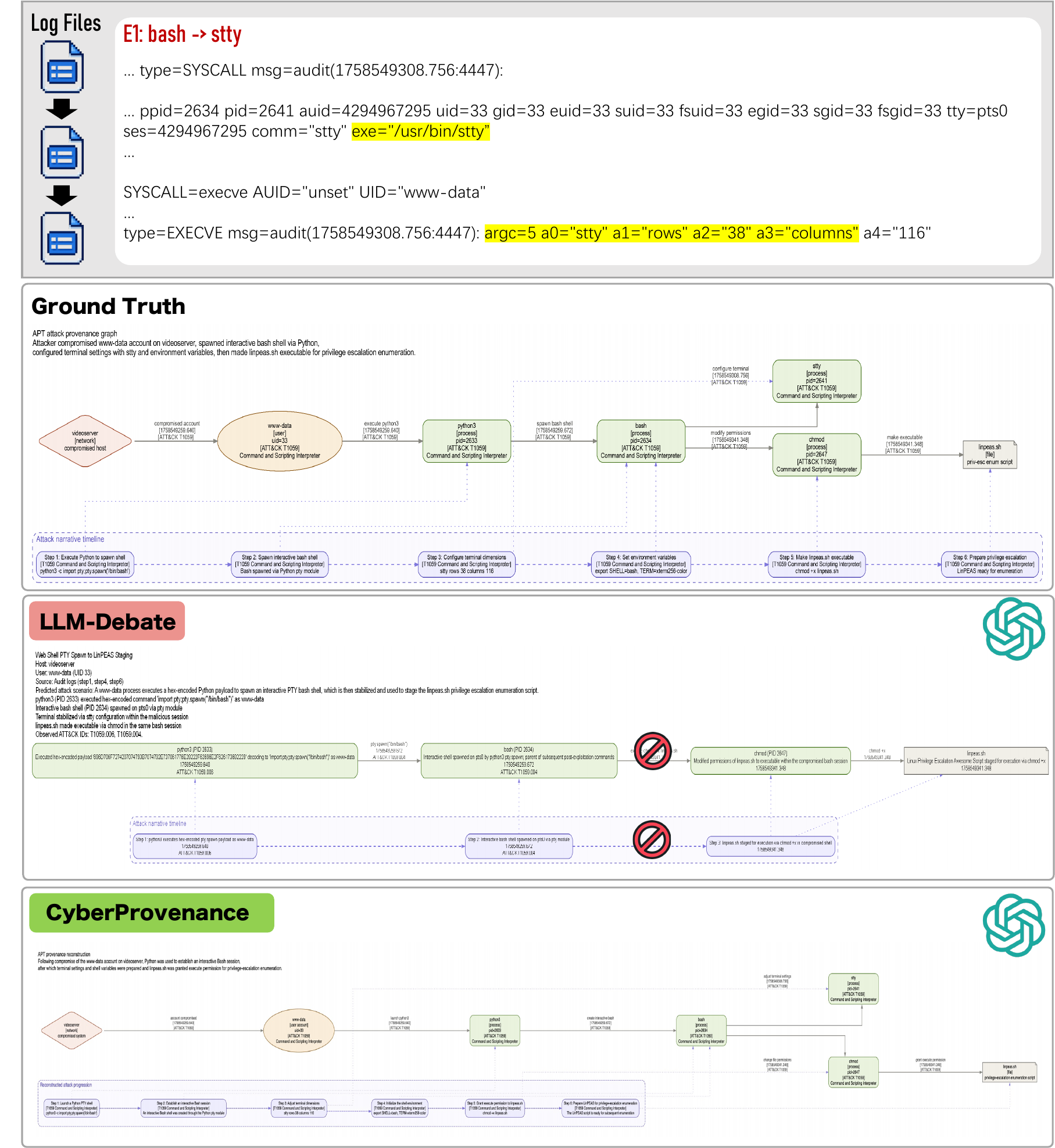}
    \caption{
Case study comparing provenance results.
LLM-Debate omits the \texttt{bash} $\rightarrow$ \texttt{stty} dependency, although the yellow-highlighted audit evidence explicitly records \texttt{/usr/bin/stty} and \texttt{stty rows 38 columns 116}.
CyberProvenance correctly aligns this evidence with the corresponding entity relation and attack step.
}
    \label{fig:case21}
\end{figure*}

\begin{figure*}[h]
    \centering
    \includegraphics[width=\linewidth]{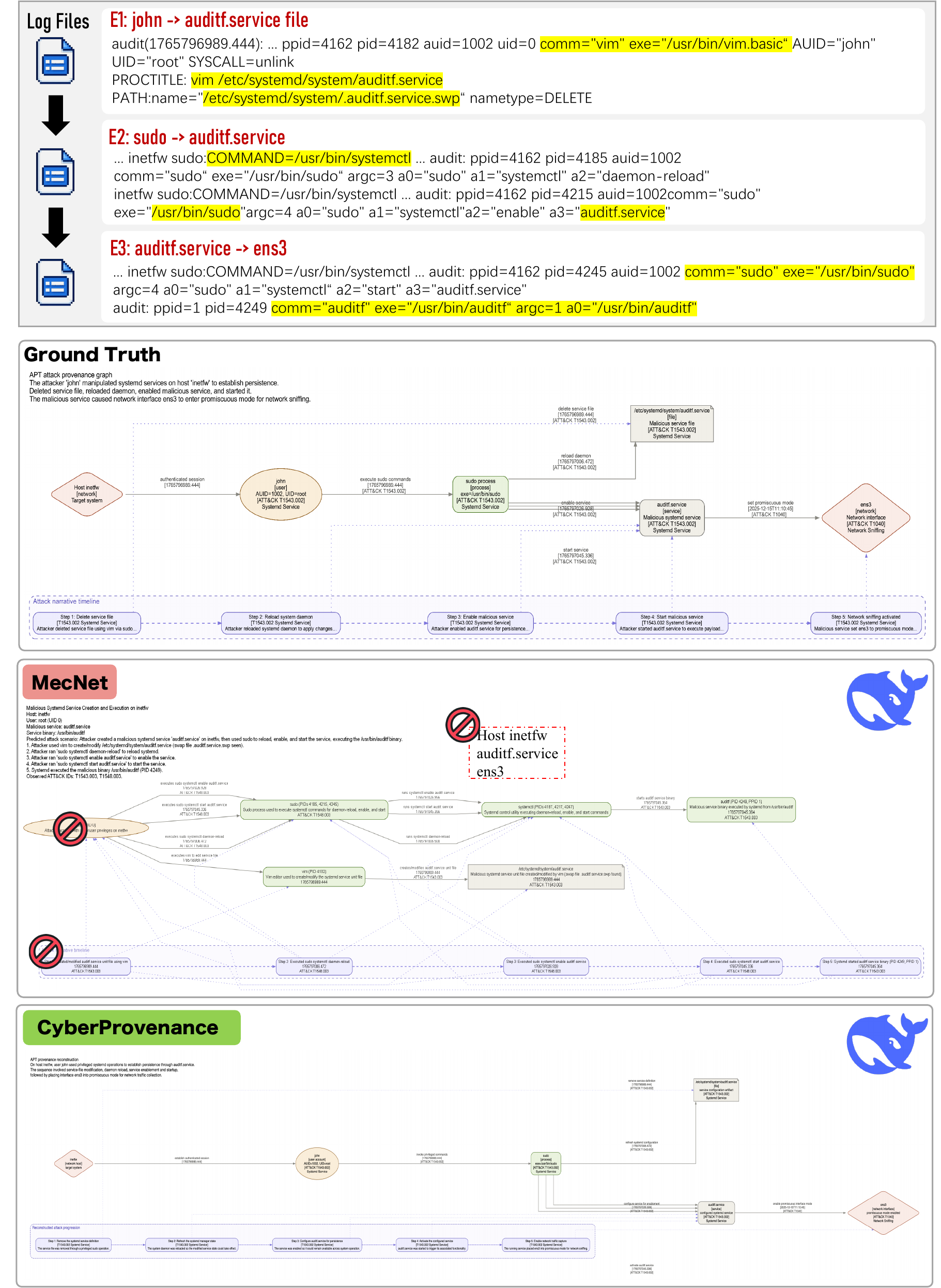}
    \caption{
Case study of attack-chain provenance.
MacNet fails to preserve three key evidence-supported relations, including the modification of \texttt{auditf.service}, its privileged activation through \texttt{sudo}, and the subsequent execution of the malicious service, while also omitting important host, service, and network entities.
As highlighted in yellow, the logs explicitly record \texttt{vim /etc/systemd/system/auditf.service}, \texttt{systemctl enable auditf.service}, and the execution of \texttt{/usr/bin/auditf}.
CyberProvenance retains these dependencies and reconstructs a more complete attack progression consistent with the observed audit evidence.
}
    \label{fig:case21}
\end{figure*}

\begin{figure*}[h]
    \centering
    \includegraphics[width=\linewidth]{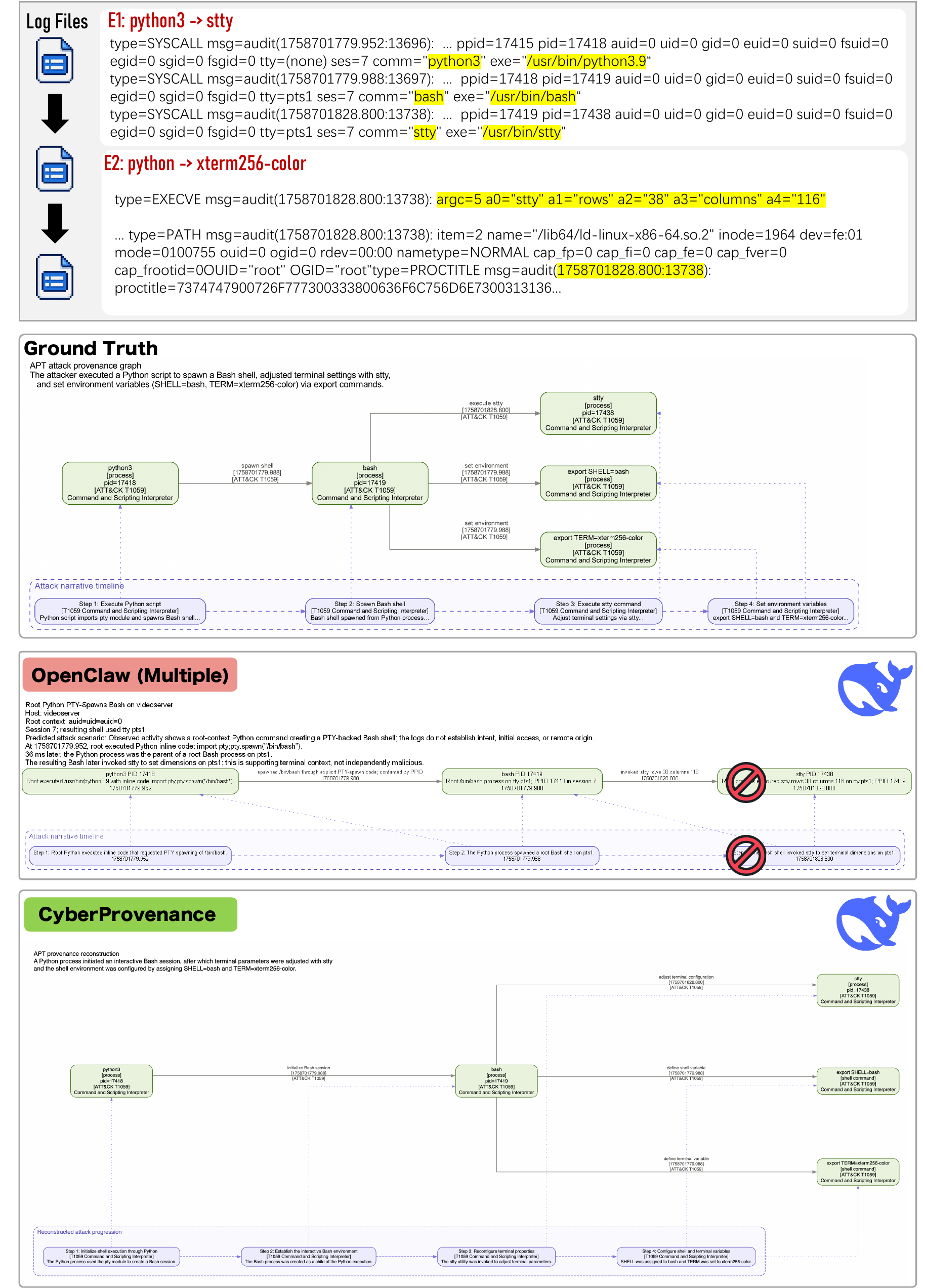}
    \caption{
Case study of attack-chain provenance.
OpenClaw fails to faithfully preserve the process dependencies associated with terminal configuration.
As highlighted in yellow, the logs explicitly record \texttt{python3} spawning \texttt{bash}, followed by the execution of \texttt{/usr/bin/stty} with \texttt{rows 38 columns 116}.
CyberProvenance correctly reconstructs these evidence-supported process relations and the corresponding attack-step progression.
}
    \label{fig:case21}
\end{figure*}

\begin{figure*}[h]
    \centering
    \includegraphics[width=0.88\linewidth]{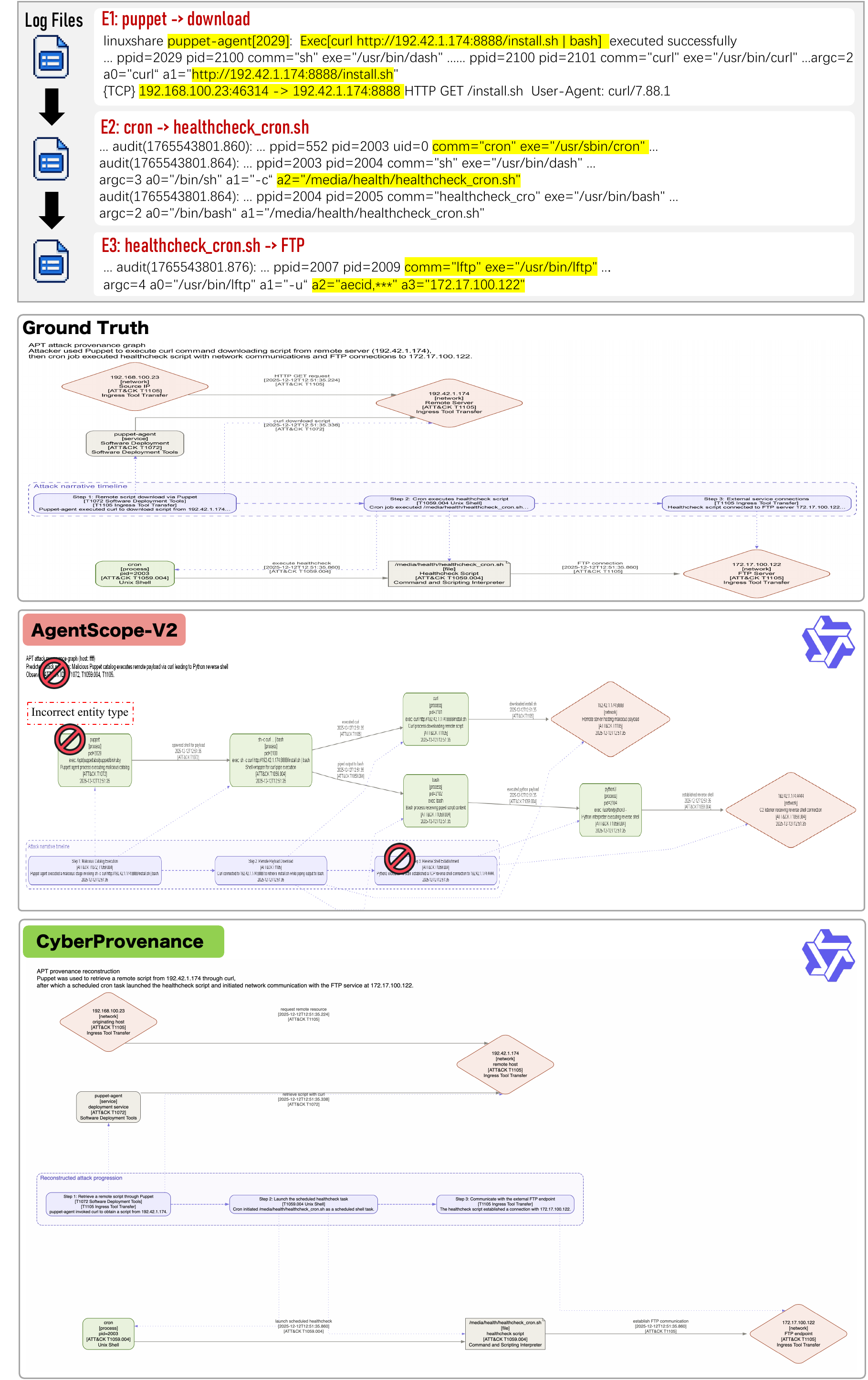}
    \caption{
Case study of attack-chain provenance.
AgentScope-V2 misclassifies a key entity and fails to faithfully align the reconstructed steps with the observed execution chain.
As highlighted in yellow, the logs explicitly record the Puppet-triggered \texttt{curl} download, the subsequent execution of \texttt{healthcheck\_cron.sh} by \texttt{cron}, and the final \texttt{lftp} connection to \texttt{172.17.100.122}.
CyberProvenance correctly reconstructs these evidence-supported entities and attack-step dependencies.
}
    \label{fig:case21}
\end{figure*}

\begin{figure*}[h]
    \centering
    \includegraphics[width=0.88\linewidth]{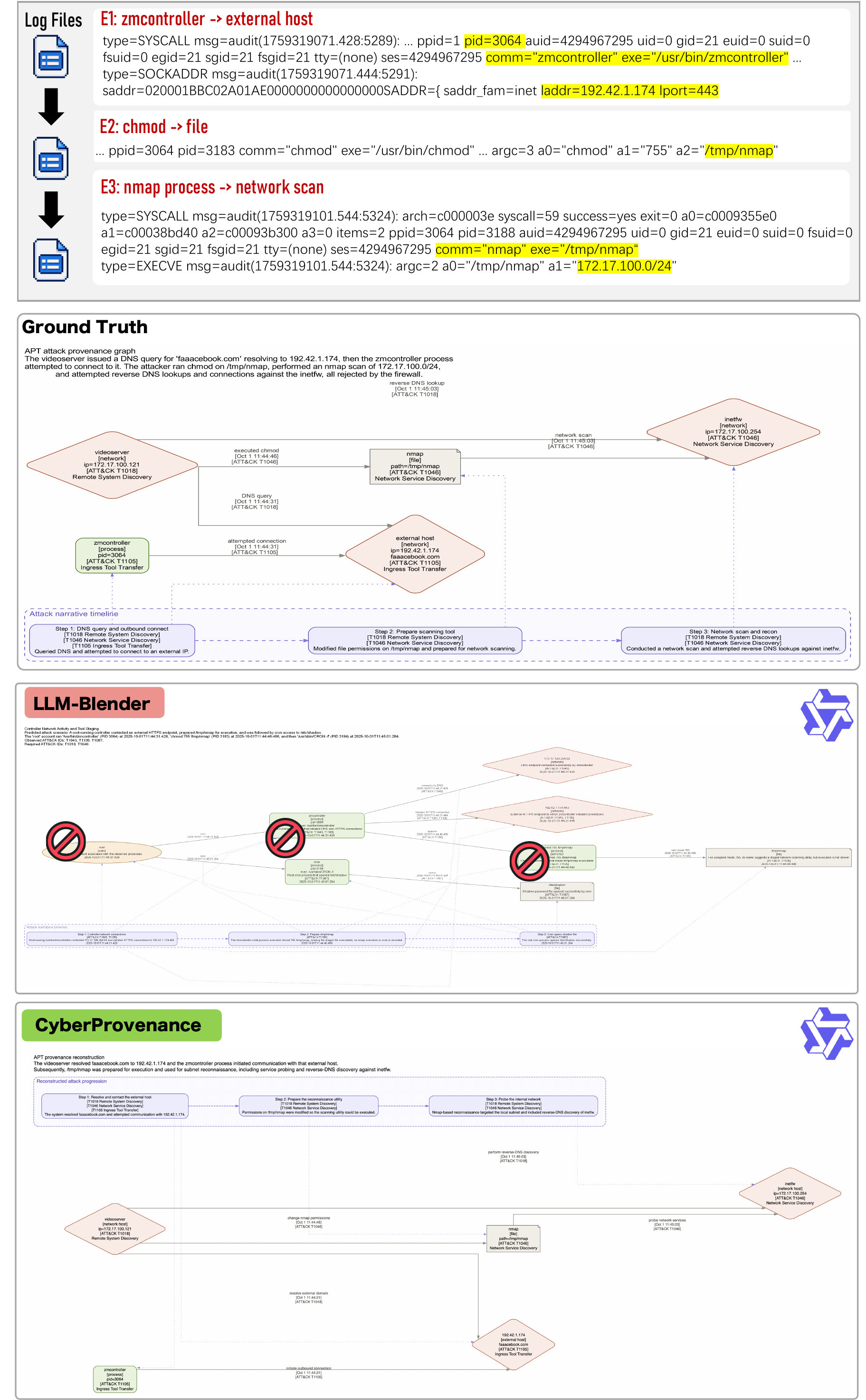}
    \caption{
Case study of attack-chain provenance.
LLM-Blender misses several key evidence-supported relations and fails to faithfully preserve the reconnaissance chain.
As highlighted in yellow, the logs explicitly record the \texttt{zmcontroller} connection to \texttt{192.42.1.174:443}, the \texttt{chmod 755 /tmp/nmap} operation, and the subsequent \texttt{nmap} scan of \texttt{172.17.100.0/24}.
CyberProvenance correctly reconstructs these observed entities, relations, and attack-step dependencies.
}
    \label{fig:case21}
\end{figure*}

\vspace{0pt}
\begin{table*}[t]
\centering
\caption{All attack behaviors in CyberClear. The 450 instances yield 89 distinct technique sets: 46 single-technique and 43 composite behaviors.}
\label{tab:attack_behaviors}
\begingroup
\fontsize{5.4}{6.0}\selectfont
\setlength{\tabcolsep}{0pt}
\renewcommand{\arraystretch}{0.90}
\setlength{\emergencystretch}{1em}
\begin{minipage}[t]{0.321\textwidth}
\vspace{0pt}
\begin{tabularx}{\linewidth}{@{}>{\raggedright\arraybackslash}X@{\hspace{2.8pt}}>{\raggedright\arraybackslash}p{1.6cm}@{}}
\toprule
\textbf{Attack behavior} & \textbf{ATT\&CK ID(s)} \\
\midrule
\textbf{B01}\enspace Abuse Elevation Control Mechanism: Sudo and Sudo Caching & \mbox{\textsf{T1548.003}} \\[0.3pt]
\textbf{B02}\enspace Account Access Removal & \mbox{\textsf{T1531}} \\[0.3pt]
\textbf{B03}\enspace Account Discovery: Local Account & \mbox{\textsf{T1087.001}} \\[0.3pt]
\textbf{B04}\enspace Active Scanning: Vulnerability Scanning & \mbox{\textsf{T1595.002}} \\[0.3pt]
\textbf{B05}\enspace Active Scanning: Wordlist Scanning & \mbox{\textsf{T1595.003}} \\[0.3pt]
\textbf{B06}\enspace Application Window Discovery & \mbox{\textsf{T1010}} \\[0.3pt]
\textbf{B07}\enspace Boot or Logon Autostart Execution & \mbox{\textsf{T1547}} \\[0.3pt]
\textbf{B08}\enspace Clipboard Data & \mbox{\textsf{T1115}} \\[0.3pt]
\textbf{B09}\enspace Command and Scripting Interpreter & \mbox{\textsf{T1059}} \\[0.3pt]
\textbf{B10}\enspace Command and Scripting Interpreter: Unix Shell & \mbox{\textsf{T1059.004}} \\[0.3pt]
\textbf{B11}\enspace Create or Modify System Process: Systemd Service & \mbox{\textsf{T1543.002}} \\[0.3pt]
\textbf{B12}\enspace Data Destruction & \mbox{\textsf{T1485}} \\[0.3pt]
\textbf{B13}\enspace Data from Information Repositories & \mbox{\textsf{T1213}} \\[0.3pt]
\textbf{B14}\enspace Data from Local System & \mbox{\textsf{T1005}} \\[0.3pt]
\textbf{B15}\enspace Data from Network Shared Drive & \mbox{\textsf{T1039}} \\[0.3pt]
\textbf{B16}\enspace Exfiltration Over C2 Channel & \mbox{\textsf{T1041}} \\[0.3pt]
\textbf{B17}\enspace Exploitation for Privilege Escalation & \mbox{\textsf{T1068}} \\[0.3pt]
\textbf{B18}\enspace File and Directory Discovery & \mbox{\textsf{T1083}} \\[0.3pt]
\textbf{B19}\enspace File and Directory Permissions Modification: Linux and Mac Permissions & \mbox{\textsf{T1222.002}} \\[0.3pt]
\textbf{B20}\enspace Gather Victim Host Information: Software & \mbox{\textsf{T1592.002}} \\[0.3pt]
\textbf{B21}\enspace Hide Artifacts: Hidden Files and Directories & \mbox{\textsf{T1564.001}} \\[0.3pt]
\textbf{B22}\enspace Indicator Removal: Clear Command History & \mbox{\textsf{T1070.003}} \\[0.3pt]
\textbf{B23}\enspace Indicator Removal: File Deletion & \mbox{\textsf{T1070.004}} \\[0.3pt]
\textbf{B24}\enspace Ingress Tool Transfer & \mbox{\textsf{T1105}} \\[0.3pt]
\textbf{B25}\enspace Inhibit System Recovery & \mbox{\textsf{T1490}} \\[0.3pt]
\textbf{B26}\enspace Network Boundary Bridging & \mbox{\textsf{T1599}} \\[0.3pt]
\textbf{B27}\enspace Network Sniffing & \mbox{\textsf{T1040}} \\[0.3pt]
\textbf{B28}\enspace OS Credential Dumping & \mbox{\textsf{T1003}} \\[0.3pt]
\textbf{B29}\enspace OS Credential Dumping: /etc/passwd and /etc/shadow & \mbox{\textsf{T1003.008}} \\[0.3pt]
\textbf{B30}\enspace Password Policy Discovery & \mbox{\textsf{T1201}} \\[0.3pt]
\textbf{B31}\enspace Peripheral Device Discovery & \mbox{\textsf{T1120}} \\[0.3pt]
\textbf{B32}\enspace Remote Services & \mbox{\textsf{T1021}} \\[0.3pt]
\textbf{B33}\enspace Remote Services: SSH & \mbox{\textsf{T1021.004}} \\[0.3pt]
\textbf{B34}\enspace Remote System Discovery & \mbox{\textsf{T1018}} \\[0.3pt]
\textbf{B35}\enspace Scheduled Task/Job & \mbox{\textsf{T1053}} \\[0.3pt]
\textbf{B36}\enspace Scheduled Task/Job: Cron & \mbox{\textsf{T1053.003}} \\[0.3pt]
\textbf{B37}\enspace Service Stop & \mbox{\textsf{T1489}} \\[0.3pt]
\textbf{B38}\enspace Software Deployment Tools & \mbox{\textsf{T1072}} \\[0.3pt]
\textbf{B39}\enspace System Information Discovery & \mbox{\textsf{T1082}} \\[0.3pt]
\textbf{B40}\enspace System Network Configuration Discovery & \mbox{\textsf{T1016}} \\[0.3pt]
\textbf{B41}\enspace System Network Connections Discovery & \mbox{\textsf{T1049}} \\[0.3pt]
\textbf{B42}\enspace System Owner/User Discovery & \mbox{\textsf{T1033}} \\[0.3pt]
\textbf{B43}\enspace System Time Discovery & \mbox{\textsf{T1124}} \\[0.3pt]
\textbf{B44}\enspace Use Alternate Authentication Material: Application Access Token & \mbox{\textsf{T1550.001}} \\[0.3pt]
\textbf{B45}\enspace Valid Accounts: Local Accounts & \mbox{\textsf{T1078.003}} \\[0.3pt]
\textbf{B46}\enspace Virtualization/Sandbox Evasion: System Checks & \mbox{\textsf{T1497.001}} \\[0.3pt]
\textbf{B47}\enspace OS Credential Dumping: /etc/passwd and /etc/shadow + Exfiltration Over C2 Channel & \mbox{\textsf{T1003.008}}\newline \mbox{\textsf{+T1041}} \\[0.3pt]
\textbf{B48}\enspace Masquerading: Match Legitimate Resource Name or Location + Traffic Signaling: Port Knocking & \mbox{\textsf{T1036.005}}\newline \mbox{\textsf{+T1205.001}} \\[0.3pt]
\textbf{B49}\enspace Masquerading: Match Legitimate Resource Name or Location + Hide Artifacts: Hidden Files and Directories & \mbox{\textsf{T1036.005}}\newline \mbox{\textsf{+T1564.001}} \\[0.3pt]
\textbf{B50}\enspace Network Sniffing + Steal Application Access Token & \mbox{\textsf{T1040}}\newline \mbox{\textsf{+T1528}} \\[0.3pt]
\textbf{B51}\enspace Command and Scripting Interpreter + Hijack Execution Flow & \mbox{\textsf{T1059}}\newline \mbox{\textsf{+T1574}} \\[0.3pt]
\textbf{B52}\enspace Command and Scripting Interpreter: Unix Shell + Remote Access Tools & \mbox{\textsf{T1059.004}}\newline \mbox{\textsf{+T1219}} \\[0.3pt]
\textbf{B53}\enspace Command and Scripting Interpreter: Python + Exploit Public-Facing Application & \mbox{\textsf{T1059.006}}\newline \mbox{\textsf{+T1190}} \\[0.3pt]
\textbf{B54}\enspace Data Staged: Local Data Staging + Archive Collected Data: Archive via Utility & \mbox{\textsf{T1074.001}}\newline \mbox{\textsf{+T1560.001}} \\[0.3pt]
\textbf{B55}\enspace Account Manipulation: SSH Authorized Keys + Ingress Tool Transfer & \mbox{\textsf{T1098.004}}\newline \mbox{\textsf{+T1105}} \\[0.3pt]
\textbf{B56}\enspace Ingress Tool Transfer + Create Account: Local Account & \mbox{\textsf{T1105}}\newline \mbox{\textsf{+T1136.001}} \\[0.3pt]
\bottomrule
\end{tabularx}
\end{minipage}
\hfill
\begin{minipage}[t]{0.321\textwidth}
\vspace{0pt}
\begin{tabularx}{\linewidth}{@{}>{\raggedright\arraybackslash}X@{\hspace{2.8pt}}>{\raggedright\arraybackslash}p{1.6cm}@{}}
\toprule
\textbf{Attack behavior} & \textbf{ATT\&CK ID(s)} \\
\midrule
\textbf{B57}\enspace Ingress Tool Transfer + Data Encrypted for Impact & \mbox{\textsf{T1105}}\newline \mbox{\textsf{+T1486}} \\[0.3pt]
\textbf{B58}\enspace Ingress Tool Transfer + Modify Authentication Process: Pluggable Authentication Modules & \mbox{\textsf{T1105}}\newline \mbox{\textsf{+T1556.003}} \\[0.3pt]
\textbf{B59}\enspace Data Encrypted for Impact + Data Manipulation: Stored Data Manipulation & \mbox{\textsf{T1486}}\newline \mbox{\textsf{+T1565.001}} \\[0.3pt]
\textbf{B60}\enspace Event Triggered Execution + Hijack Execution Flow: Path Interception by Search Order Hijacking & \mbox{\textsf{T1546}}\newline \mbox{\textsf{+T1574.008}} \\[0.3pt]
\textbf{B61}\enspace Boot or Logon Autostart Execution + System Services & \mbox{\textsf{T1547}}\newline \mbox{\textsf{+T1569}} \\[0.3pt]
\textbf{B62}\enspace Gather Victim Network Information: DNS + Gather Victim Org Information & \mbox{\textsf{T1590.002}}\newline \mbox{\textsf{+T1591}} \\[0.3pt]
\textbf{B63}\enspace Gather Victim Host Information: Software + Active Scanning & \mbox{\textsf{T1592.002}}\newline \mbox{\textsf{+T1595}} \\[0.3pt]
\textbf{B64}\enspace Rootkit + Hide Artifacts + Hijack Execution Flow & \mbox{\textsf{T1014}}\newline \mbox{\textsf{+T1564}}\newline \mbox{\textsf{+T1574}} \\[0.3pt]
\textbf{B65}\enspace Remote System Discovery + Network Service Discovery + Ingress Tool Transfer & \mbox{\textsf{T1018}}\newline \mbox{\textsf{+T1046}}\newline \mbox{\textsf{+T1105}} \\[0.3pt]
\textbf{B66}\enspace Masquerading: Match Legitimate Resource Name or Location + Application Layer Protocol: Web Protocols + Traffic Signaling: Port Knocking & \mbox{\textsf{T1036.005}}\newline \mbox{\textsf{+T1071.001}}\newline \mbox{\textsf{+T1205.001}} \\[0.3pt]
\textbf{B67}\enspace Scheduled Task/Job: Cron + Application Layer Protocol: Web Protocols + Ingress Tool Transfer & \mbox{\textsf{T1053.003}}\newline \mbox{\textsf{+T1071.001}}\newline \mbox{\textsf{+T1105}} \\[0.3pt]
\textbf{B68}\enspace Command and Scripting Interpreter: Unix Shell + Software Deployment Tools + Ingress Tool Transfer & \mbox{\textsf{T1059.004}}\newline \mbox{\textsf{+T1072}}\newline \mbox{\textsf{+T1105}} \\[0.3pt]
\textbf{B69}\enspace Command and Scripting Interpreter: Unix Shell + Non-Application Layer Protocol + Exploit Public-Facing Application & \mbox{\textsf{T1059.004}}\newline \mbox{\textsf{+T1095}}\newline \mbox{\textsf{+T1190}} \\[0.3pt]
\textbf{B70}\enspace Valid Accounts: Domain Accounts + Brute Force: Password Guessing + External Remote Services & \mbox{\textsf{T1078.002}}\newline \mbox{\textsf{+T1110.001}}\newline \mbox{\textsf{+T1133}} \\[0.3pt]
\textbf{B71}\enspace Valid Accounts: Local Accounts + Ingress Tool Transfer + Traffic Signaling: Port Knocking & \mbox{\textsf{T1078.003}}\newline \mbox{\textsf{+T1105}}\newline \mbox{\textsf{+T1205.001}} \\[0.3pt]
\textbf{B72}\enspace Account Manipulation: SSH Authorized Keys + Ingress Tool Transfer + System Binary Proxy Execution & \mbox{\textsf{T1098.004}}\newline \mbox{\textsf{+T1105}}\newline \mbox{\textsf{+T1218}} \\[0.3pt]
\textbf{B73}\enspace Ingress Tool Transfer + Create Account: Local Account + System Binary Proxy Execution & \mbox{\textsf{T1105}}\newline \mbox{\textsf{+T1136.001}}\newline \mbox{\textsf{+T1218}} \\[0.3pt]
\textbf{B74}\enspace Ingress Tool Transfer + User Execution: Malicious Link + Phishing: Spearphishing Voice & \mbox{\textsf{T1105}}\newline \mbox{\textsf{+T1204.001}}\newline \mbox{\textsf{+T1566.004}} \\[0.3pt]
\textbf{B75}\enspace Ingress Tool Transfer + System Binary Proxy Execution + Modify Authentication Process: Pluggable Authentication Modules & \mbox{\textsf{T1105}}\newline \mbox{\textsf{+T1218}}\newline \mbox{\textsf{+T1556.003}} \\[0.3pt]
\textbf{B76}\enspace Ingress Tool Transfer + Remote Access Tools + Hide Artifacts: Hidden Files and Directories & \mbox{\textsf{T1105}}\newline \mbox{\textsf{+T1219}}\newline \mbox{\textsf{+T1564.001}} \\[0.3pt]
\textbf{B77}\enspace Remote Services: SMB/Windows Admin Shares + Exfiltration Over C2 Channel + Command and Scripting Interpreter: PowerShell + Application Layer Protocol: Web Protocols & \mbox{\textsf{T1021.002}}\newline \mbox{\textsf{+T1041}}\newline \mbox{\textsf{+T1059.001}}\newline \mbox{\textsf{+T1071.001}} \\[0.3pt]
\textbf{B78}\enspace Scheduled Task/Job: Cron + Exploitation of Remote Services + Implant Internal Image + Deploy Container & \mbox{\textsf{T1053.003}}\newline \mbox{\textsf{+T1210}}\newline \mbox{\textsf{+T1525}}\newline \mbox{\textsf{+T1610}} \\[0.3pt]
\textbf{B79}\enspace Input Capture: Keylogging + Input Capture: Credential API Hooking + Clipboard Data + Software Extensions & \mbox{\textsf{T1056.001}}\newline \mbox{\textsf{+T1056.004}}\newline \mbox{\textsf{+T1115}}\newline \mbox{\textsf{+T1176}} \\[0.3pt]
\textbf{B80}\enspace Exfiltration Over C2 Channel + Command and Scripting Interpreter: PowerShell + Indicator Removal: File Deletion + Phishing: Spearphishing Attachment + Acquire Infrastructure: Server & \mbox{\textsf{T1041}}\newline \mbox{\textsf{+T1059.001}}\newline \mbox{\textsf{+T1070.004}}\newline \mbox{\textsf{+T1566.001}}\newline \mbox{\textsf{+T1583.004}} \\[0.3pt]
\textbf{B81}\enspace Exfiltration Over C2 Channel + Exploit Public-Facing Application + Create or Modify System Process: Windows Service + Phishing: Spearphishing Attachment + Acquire Infrastructure: Server & \mbox{\textsf{T1041}}\newline \mbox{\textsf{+T1190}}\newline \mbox{\textsf{+T1543.003}}\newline \mbox{\textsf{+T1566.001}}\newline \mbox{\textsf{+T1583.004}} \\[0.3pt]
\bottomrule
\end{tabularx}
\end{minipage}
\hfill
\begin{minipage}[t]{0.321\textwidth}
\vspace{0pt}
\begin{tabularx}{\linewidth}{@{}>{\raggedright\arraybackslash}X@{\hspace{2.8pt}}>{\raggedright\arraybackslash}p{1.6cm}@{}}
\toprule
\textbf{Attack behavior} & \textbf{ATT\&CK ID(s)} \\
\midrule
\textbf{B82}\enspace OS Credential Dumping: LSASS Memory + OS Credential Dumping: Cached Domain Credentials + Input Capture: Keylogging + Email Collection: Remote Email Collection + Create or Modify System Process: Windows Service + Phishing: Spearphishing Attachment & \mbox{\textsf{T1003.001}}\newline \mbox{\textsf{+T1003.005}}\newline \mbox{\textsf{+T1056.001}}\newline \mbox{\textsf{+T1114.002}}\newline \mbox{\textsf{+T1543.003}}\newline \mbox{\textsf{+T1566.001}} \\[0.3pt]
\textbf{B83}\enspace OS Credential Dumping: LSASS Memory + OS Credential Dumping: Cached Domain Credentials + Input Capture: Keylogging + Command and Scripting Interpreter + Email Collection: Local Email Collection + Email Collection: Remote Email Collection + Phishing: Spearphishing Attachment & \mbox{\textsf{T1003.001}}\newline \mbox{\textsf{+T1003.005}}\newline \mbox{\textsf{+T1056.001}}\newline \mbox{\textsf{+T1059}}\newline \mbox{\textsf{+T1114.001}}\newline \mbox{\textsf{+T1114.002}}\newline \mbox{\textsf{+T1566.001}} \\[0.3pt]
\textbf{B84}\enspace OS Credential Dumping: LSASS Memory + OS Credential Dumping: Cached Domain Credentials + Input Capture: Keylogging + Email Collection: Local Email Collection + Email Collection: Remote Email Collection + Create or Modify System Process: Windows Service + Phishing: Spearphishing Attachment & \mbox{\textsf{T1003.001}}\newline \mbox{\textsf{+T1003.005}}\newline \mbox{\textsf{+T1056.001}}\newline \mbox{\textsf{+T1114.001}}\newline \mbox{\textsf{+T1114.002}}\newline \mbox{\textsf{+T1543.003}}\newline \mbox{\textsf{+T1566.001}} \\[0.3pt]
\textbf{B85}\enspace OS Credential Dumping: LSASS Memory + OS Credential Dumping: Cached Domain Credentials + Remote Services + Input Capture: Keylogging + Email Collection: Local Email Collection + Email Collection: Remote Email Collection + Create or Modify System Process: Windows Service + Phishing: Spearphishing Attachment & \mbox{\textsf{T1003.001}}\newline \mbox{\textsf{+T1003.005}}\newline \mbox{\textsf{+T1021}}\newline \mbox{\textsf{+T1056.001}}\newline \mbox{\textsf{+T1114.001}}\newline \mbox{\textsf{+T1114.002}}\newline \mbox{\textsf{+T1543.003}}\newline \mbox{\textsf{+T1566.001}} \\[0.3pt]
\textbf{B86}\enspace Remote Services: Windows Remote Management + Exfiltration Over C2 Channel + Command and Scripting Interpreter: PowerShell + Web Service: Bidirectional Communication + Ingress Tool Transfer + Exfiltration Over Web Service: Exfiltration to Cloud Storage + Lateral Tool Transfer + Acquire Infrastructure: Server & \mbox{\textsf{T1021.006}}\newline \mbox{\textsf{+T1041}}\newline \mbox{\textsf{+T1059.001}}\newline \mbox{\textsf{+T1102.002}}\newline \mbox{\textsf{+T1105}}\newline \mbox{\textsf{+T1567.002}}\newline \mbox{\textsf{+T1570}}\newline \mbox{\textsf{+T1583.004}} \\[0.3pt]
\textbf{B87}\enspace OS Credential Dumping: LSASS Memory + OS Credential Dumping: Cached Domain Credentials + Remote Services + Input Capture: Keylogging + Command and Scripting Interpreter + Application Layer Protocol: Web Protocols + Email Collection: Local Email Collection + Email Collection: Remote Email Collection + Phishing: Spearphishing Attachment & \mbox{\textsf{T1003.001}}\newline \mbox{\textsf{+T1003.005}}\newline \mbox{\textsf{+T1021}}\newline \mbox{\textsf{+T1056.001}}\newline \mbox{\textsf{+T1059}}\newline \mbox{\textsf{+T1071.001}}\newline \mbox{\textsf{+T1114.001}}\newline \mbox{\textsf{+T1114.002}}\newline \mbox{\textsf{+T1566.001}} \\[0.3pt]
\textbf{B88}\enspace Exfiltration Over C2 Channel + Command and Scripting Interpreter: PowerShell + Application Layer Protocol: Web Protocols + Account Discovery: Local Account + Drive-by Compromise + System Binary Proxy Execution: Mshta + System Binary Proxy Execution: Msiexec + Create or Modify System Process: Windows Service + Boot or Logon Autostart Execution: Registry Run Keys / Startup Folder + Acquire Infrastructure: Web Services & \mbox{\textsf{T1041}}\newline \mbox{\textsf{+T1059.001}}\newline \mbox{\textsf{+T1071.001}}\newline \mbox{\textsf{+T1087.001}}\newline \mbox{\textsf{+T1189}}\newline \mbox{\textsf{+T1218.005}}\newline \mbox{\textsf{+T1218.007}}\newline \mbox{\textsf{+T1543.003}}\newline \mbox{\textsf{+T1547.001}}\newline \mbox{\textsf{+T1583.006}} \\[0.3pt]
\textbf{B89}\enspace System Service Discovery + System Network Configuration Discovery + System Owner/User Discovery + System Network Connections Discovery + Process Discovery + Permission Groups Discovery + System Information Discovery + File and Directory Discovery + Account Discovery + Password Policy Discovery + Software Discovery + System Location Discovery + Group Policy Discovery & \mbox{\textsf{T1007}}\newline \mbox{\textsf{+T1016}}\newline \mbox{\textsf{+T1033}}\newline \mbox{\textsf{+T1049}}\newline \mbox{\textsf{+T1057}}\newline \mbox{\textsf{+T1069}}\newline \mbox{\textsf{+T1082}}\newline \mbox{\textsf{+T1083}}\newline \mbox{\textsf{+T1087}}\newline \mbox{\textsf{+T1201}}\newline \mbox{\textsf{+T1518}}\newline \mbox{\textsf{+T1614}}\newline \mbox{\textsf{+T1615}} \\[0.3pt]
\bottomrule
\end{tabularx}
\end{minipage}
\par\vspace{4pt}

\endgroup
\end{table*}

\begin{table*}[t]
\centering

\caption{Attack behavior coverage of different agent frameworks under GPT-5.6-Terra.
A checkmark indicates that the framework successfully handles the corresponding attack behavior,
while a cross indicates failure.}
\vspace{6pt}
\label{tab:gpt_behavior_coverage}
\fontsize{5.4}{6.0}\selectfont
\setlength{\belowcaptionskip}{2pt}
\setlength{\tabcolsep}{2.2pt}
\renewcommand{\arraystretch}{0.68}

\resizebox{\textwidth}{!}{
\begin{tabular}{
@{}
c
ccccccccccc
@{}
}

\toprule

\textbf{Behavior}
& \textbf{OpenClaw (Single)}
& \textbf{OpenClaw (Multiple)}
& \textbf{AgentVerse}
& \textbf{LLM-Debate}
& \textbf{MultiPersona}
& \textbf{LLM-Blender}
& \textbf{DyLAN}
& \textbf{MacNet}
& \textbf{CAMEL}
& \textbf{AgentScope-V2}
& \textbf{CyberProvenance}
\\

\midrule

B01 & \cmark & \cmark & \cmark & \cmark & \cmark & \cmark & \cmark & \cmark & \cmark & \cmark & \cmark \\
B02 & \cmark & \cmark & \cmark & \cmark & \cmark & \cmark & \cmark & \cmark & \cmark & \cmark & \cmark \\
B03 & \cmark & \cmark & \cmark & \cmark & \cmark & \cmark & \cmark & \cmark & \cmark & \cmark & \cmark \\
B04 & \cmark & \cmark & \cmark & \cmark & \cmark & \cmark & \cmark & \cmark & \cmark & \cmark & \cmark \\
B05 & \cmark & \cmark & \cmark & \cmark & \cmark & \cmark & \cmark & \cmark & \cmark & \cmark & \cmark \\
B06 & \cmark & \cmark & \cmark & \cmark & \cmark & \cmark & \cmark & \cmark & \cmark & \cmark & \cmark \\
B07 & \cmark & \cmark & \cmark & \cmark & \cmark & \cmark & \cmark & \cmark & \cmark & \cmark & \cmark \\
B08 & \cmark & \cmark & \cmark & \cmark & \cmark & \cmark & \cmark & \cmark & \cmark & \cmark & \cmark \\
B09 & \cmark & \cmark & \cmark & \cmark & \cmark & \cmark & \cmark & \cmark & \cmark & \cmark & \cmark \\
B10 & \cmark & \cmark & \cmark & \cmark & \cmark & \cmark & \cmark & \cmark & \cmark & \cmark & \cmark \\
B11 & \cmark & \cmark & \cmark & \cmark & \cmark & \cmark & \cmark & \cmark & \cmark & \cmark & \cmark \\
B12 & \cmark & \cmark & \cmark & \cmark & \cmark & \cmark & \cmark & \cmark & \cmark & \cmark & \cmark \\
B13 & \cmark & \cmark & \cmark & \cmark & \cmark & \cmark & \cmark & \cmark & \cmark & \cmark & \cmark \\
B14 & \cmark & \cmark & \cmark & \cmark & \cmark & \cmark & \cmark & \cmark & \cmark & \cmark & \cmark \\
B15 & \cmark & \cmark & \cmark & \cmark & \cmark & \cmark & \cmark & \cmark & \cmark & \cmark & \cmark \\
B16 & \cmark & \cmark & \cmark & \cmark & \cmark & \cmark & \cmark & \cmark & \cmark & \cmark & \cmark \\

B17 & \xmark & \cmark & \cmark & \xmark & \cmark & \xmark & \xmark & \xmark & \cmark & \cmark & \cmark \\
B18 & \cmark & \cmark & \cmark & \cmark & \cmark & \cmark & \cmark & \cmark & \cmark & \cmark & \cmark \\
B19 & \cmark & \cmark & \cmark & \cmark & \cmark & \cmark & \cmark & \cmark & \cmark & \cmark & \cmark \\
B20 & \cmark & \cmark & \cmark & \cmark & \cmark & \cmark & \cmark & \cmark & \cmark & \cmark & \cmark \\
B21 & \cmark & \cmark & \cmark & \cmark & \cmark & \cmark & \cmark & \cmark & \cmark & \cmark & \cmark \\
B22 & \cmark & \cmark & \cmark & \cmark & \cmark & \cmark & \cmark & \cmark & \cmark & \cmark & \cmark \\
B23 & \cmark & \cmark & \cmark & \cmark & \cmark & \cmark & \cmark & \cmark & \cmark & \cmark & \cmark \\
B24 & \cmark & \cmark & \cmark & \cmark & \cmark & \cmark & \cmark & \cmark & \cmark & \cmark & \cmark \\
B25 & \cmark & \cmark & \cmark & \cmark & \cmark & \cmark & \xmark & \cmark & \cmark & \cmark & \cmark \\
B26 & \cmark & \cmark & \cmark & \cmark & \cmark & \cmark & \cmark & \cmark & \cmark & \cmark & \cmark \\
B27 & \cmark & \cmark & \cmark & \cmark & \cmark & \cmark & \cmark & \cmark & \cmark & \cmark & \cmark \\
B28 & \xmark & \xmark & \cmark & \xmark & \xmark & \xmark & \xmark & \xmark & \xmark & \cmark & \cmark \\
B29 & \cmark & \cmark & \cmark & \cmark & \cmark & \cmark & \cmark & \cmark & \cmark & \cmark & \cmark \\
B30 & \cmark & \cmark & \cmark & \cmark & \cmark & \cmark & \cmark & \cmark & \cmark & \cmark & \cmark \\
B31 & \cmark & \cmark & \cmark & \cmark & \cmark & \cmark & \cmark & \cmark & \cmark & \cmark & \cmark \\
B32 & \cmark & \cmark & \cmark & \cmark & \cmark & \cmark & \cmark & \cmark & \cmark & \cmark & \cmark \\
B33 & \cmark & \cmark & \cmark & \cmark & \cmark & \cmark & \cmark & \cmark & \cmark & \cmark & \cmark \\
B34 & \cmark & \cmark & \cmark & \cmark & \cmark & \cmark & \cmark & \cmark & \cmark & \cmark & \cmark \\
B35 & \cmark & \cmark & \cmark & \cmark & \cmark & \cmark & \cmark & \cmark & \cmark & \cmark & \cmark \\
B36 & \cmark & \cmark & \cmark & \cmark & \cmark & \cmark & \cmark & \cmark & \cmark & \cmark & \cmark \\
B37 & \cmark & \cmark & \cmark & \cmark & \cmark & \cmark & \cmark & \cmark & \cmark & \cmark & \cmark \\
B38 & \xmark & \cmark & \cmark & \xmark & \cmark & \cmark & \cmark & \xmark & \cmark & \cmark & \cmark \\
B39 & \cmark & \cmark & \cmark & \cmark & \cmark & \cmark & \cmark & \cmark & \cmark & \cmark & \cmark \\
B40 & \cmark & \cmark & \cmark & \cmark & \cmark & \cmark & \cmark & \cmark & \cmark & \cmark & \cmark \\
B41 & \cmark & \cmark & \cmark & \cmark & \cmark & \cmark & \cmark & \cmark & \cmark & \cmark & \cmark \\
B42 & \cmark & \cmark & \cmark & \cmark & \cmark & \cmark & \cmark & \cmark & \cmark & \cmark & \cmark \\
B43 & \cmark & \cmark & \cmark & \cmark & \cmark & \cmark & \cmark & \cmark & \cmark & \cmark & \cmark \\
B44 & \xmark & \cmark & \xmark & \cmark & \cmark & \cmark & \cmark & \xmark & \cmark & \cmark & \cmark \\
B45 & \cmark & \cmark & \cmark & \cmark & \cmark & \cmark & \cmark & \cmark & \cmark & \cmark & \cmark \\
B46 & \cmark & \cmark & \cmark & \cmark & \cmark & \cmark & \cmark & \cmark & \cmark & \cmark & \cmark \\

\midrule

B47 & \cmark & \cmark & \cmark & \cmark & \cmark & \cmark & \cmark & \cmark & \cmark & \cmark & \xmark \\
B48 & \xmark & \cmark & \cmark & \xmark & \cmark & \xmark & \xmark & \xmark & \cmark & \xmark & \cmark \\
B49 & \cmark & \cmark & \cmark & \cmark & \cmark & \cmark & \xmark & \cmark & \cmark & \cmark & \cmark \\
B50 & \xmark & \cmark & \xmark & \xmark & \cmark & \cmark & \xmark & \xmark & \cmark & \cmark & \cmark \\
B51 & \cmark & \cmark & \cmark & \cmark & \cmark & \cmark & \xmark & \cmark & \cmark & \xmark & \cmark \\
B52 & \cmark & \cmark & \cmark & \cmark & \cmark & \cmark & \cmark & \cmark & \cmark & \cmark & \cmark \\
B53 & \xmark & \xmark & \cmark & \xmark & \xmark & \xmark & \xmark & \xmark & \xmark & \xmark & \cmark \\
B54 & \cmark & \cmark & \cmark & \cmark & \cmark & \cmark & \cmark & \cmark & \cmark & \xmark & \xmark \\
B55 & \cmark & \cmark & \cmark & \cmark & \cmark & \cmark & \cmark & \cmark & \cmark & \xmark & \cmark \\
B56 & \xmark & \cmark & \xmark & \xmark & \xmark & \cmark & \xmark & \cmark & \cmark & \cmark & \cmark \\
B57 & \xmark & \cmark & \cmark & \xmark & \xmark & \xmark & \xmark & \xmark & \cmark & \xmark & \cmark \\
B58 & \xmark & \xmark & \cmark & \cmark & \xmark & \cmark & \xmark & \xmark & \cmark & \xmark & \cmark \\
B59 & \xmark & \cmark & \cmark & \cmark & \cmark & \cmark & \xmark & \cmark & \cmark & \xmark & \cmark \\
B60 & \xmark & \xmark & \xmark & \xmark & \xmark & \xmark & \xmark & \xmark & \xmark & \cmark & \cmark \\
B61 & \cmark & \xmark & \cmark & \cmark & \xmark & \cmark & \cmark & \cmark & \cmark & \xmark & \cmark \\
B62 & \xmark & \cmark & \cmark & \xmark & \cmark & \xmark & \xmark & \xmark & \cmark & \xmark & \cmark \\
B63 & \cmark & \xmark & \cmark & \cmark & \cmark & \cmark & \cmark & \xmark & \cmark & \cmark & \cmark \\

B64 & \xmark & \cmark & \xmark & \xmark & \cmark & \xmark & \xmark & \xmark & \xmark & \cmark & \cmark \\
B65 & \cmark & \xmark & \xmark & \xmark & \cmark & \cmark & \cmark & \cmark & \cmark & \xmark & \cmark \\
B66 & \xmark & \cmark & \cmark & \xmark & \cmark & \cmark & \xmark & \xmark & \xmark & \cmark & \xmark \\
B67 & \cmark & \xmark & \cmark & \cmark & \cmark & \cmark & \cmark & \cmark & \xmark & \cmark & \cmark \\
B68 & \cmark & \cmark & \cmark & \cmark & \cmark & \cmark & \cmark & \cmark & \xmark & \xmark & \cmark \\
B69 & \xmark & \xmark & \xmark & \xmark & \xmark & \xmark & \xmark & \xmark & \xmark & \cmark & \cmark \\
B70 & \cmark & \cmark & \cmark & \cmark & \cmark & \cmark & \cmark & \xmark & \cmark & \xmark & \cmark \\
B71 & \xmark & \cmark & \cmark & \xmark & \xmark & \xmark & \xmark & \xmark & \cmark & \xmark & \cmark \\
B72 & \cmark & \cmark & \cmark & \xmark & \xmark & \cmark & \cmark & \cmark & \xmark & \cmark & \cmark \\
B73 & \xmark & \xmark & \xmark & \xmark & \cmark & \cmark & \xmark & \xmark & \xmark & \cmark & \xmark \\
B74 & \xmark & \cmark & \cmark & \xmark & \xmark & \cmark & \cmark & \cmark & \xmark & \cmark & \cmark \\
B75 & \xmark & \cmark & \cmark & \cmark & \xmark & \cmark & \cmark & \xmark & \cmark & \cmark & \cmark \\
B76 & \xmark & \xmark & \cmark & \xmark & \xmark & \xmark & \xmark & \xmark & \xmark & \xmark & \cmark \\

B77 & \xmark & \xmark & \xmark & \xmark & \xmark & \xmark & \cmark & \xmark & \xmark & \xmark & \xmark \\
B78 & \xmark & \cmark & \xmark & \xmark & \xmark & \cmark & \cmark & \cmark & \xmark & \cmark & \cmark \\
B79 & \xmark & \xmark & \xmark & \xmark & \xmark & \xmark & \xmark & \xmark & \cmark & \xmark & \xmark \\
B80 & \xmark & \cmark & \cmark & \xmark & \xmark & \cmark & \xmark & \xmark & \xmark & \xmark & \cmark \\
B81 & \xmark & \xmark & \cmark & \xmark & \cmark & \xmark & \xmark & \xmark & \cmark & \xmark & \xmark \\

B82 & \xmark & \xmark & \xmark & \xmark & \xmark & \xmark & \xmark & \xmark & \xmark & \xmark & \xmark \\
B83 & \xmark & \xmark & \xmark & \xmark & \xmark & \xmark & \xmark & \xmark & \xmark & \xmark & \xmark \\
B84 & \xmark & \xmark & \xmark & \xmark & \xmark & \xmark & \xmark & \xmark & \xmark & \xmark & \xmark \\
B85 & \xmark & \xmark & \xmark & \xmark & \xmark & \xmark & \xmark & \xmark & \xmark & \xmark & \xmark \\
B86 & \xmark & \xmark & \xmark & \xmark & \xmark & \xmark & \xmark & \xmark & \xmark & \xmark & \xmark \\
B87 & \xmark & \xmark & \xmark & \xmark & \xmark & \xmark & \xmark & \xmark & \xmark & \xmark & \xmark \\
B88 & \xmark & \xmark & \xmark & \xmark & \xmark & \xmark & \xmark & \xmark & \xmark & \xmark & \xmark \\
B89 & \xmark & \xmark & \xmark & \xmark & \xmark & \xmark & \xmark & \xmark & \xmark & \xmark & \xmark \\

\bottomrule

\end{tabular}
}

\vspace{-8pt}
\end{table*}

\end{document}